\documentclass[conference]{IEEEtran}
\IEEEoverridecommandlockouts
\ifCLASSINFOpdf
\else
\fi
\usepackage{hyperref} 
\usepackage{graphicx}
\usepackage[linesnumbered,lined,vlined,ruled,commentsnumbered,noend]{algorithm2e}
\usepackage{textcomp}
\usepackage{listings}
\usepackage{url}
\usepackage{xspace}
\usepackage{multirow}
\usepackage{balance}
\usepackage[most]{tcolorbox}
\tcbuselibrary{listings, breakable}
\usepackage{ctable}
\usepackage{listings}
\usepackage{color}
\usepackage{graphicx}
\usepackage{subcaption}
\usepackage[nobiblatex]{xurl}

\usepackage{amsmath}

\usepackage{amssymb}
\usepackage{amsfonts}

\usepackage{multirow}

\usepackage{algpseudocode}
\usepackage{caption}
\usepackage{subcaption}

\usepackage{cleveref}
\usepackage{acronym}
\usepackage{fancybox}
\usepackage{verbatim}
\usepackage[normalem]{ulem}
\usepackage{float}
\usepackage{newfloat}
\usepackage{mdframed}
\usepackage{enumitem}
\usepackage{booktabs}
\usepackage{pifont}
\usepackage{tabularray}
\usepackage{array}
\usepackage{colortbl}
\tcbuselibrary{breakable}

\usepackage{tikz} % 用于绘制图形
\usepackage{xcolor} % 用于颜色设置

\usepackage[export]{adjustbox} %图字体大小
\definecolor{red}{RGB}{255,0,0}
\definecolor{green}{RGB}{18,220,168}

\newcommand{\eg}{e.g.,\xspace}

\definecolor{dkgreen}{rgb}{0,0.6,0}

\SetCommentSty{mycommfont}

\newcommand{\toolname}{\textsc{KnowHow}\xspace}
\newcommand{\extractor}{\textsc{Extractor}\xspace}
\newcommand{\ladder}{\textsc{LADDER}\xspace}
\newcommand{\threatraptor}{\textsc{ThreatRaptor}\xspace}
\newcommand{\attackg}{\textsc{AttacKG}\xspace}
\newcommand{\nodlink}{\textsc{NodLink}\xspace}
\newcommand{\kairos}{\textsc{Kairos}\xspace}
\newcommand{\holmes}{\textsc{Holmes}\xspace}
\newcommand{\airtag}{\textsc{AirTag}\xspace}

\newcommand{\poirot}{\textsc{Poirot}\xspace}
\newcommand{\rapsheet}{\textsc{RapSheet}\xspace}
\newcommand{\ttpdrill}{\textsc{TTPDrill}\xspace}

\newcommand{\trec}{\textsc{TREC}\xspace}
\newcommand{\aptshield}{\textsc{APTSHIELD}\xspace}

\newcommand{\eat}[1]{}
\acrodef{wp}[WP]{Website Fingerprinting}
\acrodef{apt}[APT]{Advanced Persistent Threats}
\acrodef{lol}[LotL]{Living-Off-The-Land}
\acrodef{ids}[IDS]{Intrusion Detection System}
\acrodef{vae}[VAE]{Variational AutoEncoder}
\acrodef{re}[RE]{reconstruction error}
\acrodef{sv}[SV]{Stableness Value}
\acrodef{as}[AS]{anomaly score}
\acrodef{gas}[GAS]{graph anomaly score}
\acrodef{etw}[ETW]{Event Tracing for Windows}
\acrodef{e3}[E3]{Engagement 3}
\acrodef{e5}[E5]{Engagement 5}
\acrodef{ttp}[TTPs]{Tactics, Techniques, and Procedures}
\acrodef{hsg}[HSG]{High-level Scenario Graph}
\acrodef{nlp}[NLP]{Natural Language Processing}
\acrodef{dg}[DG]{Detection Graph}
\acrodef{poi}[POI]{Point of Interest}
\acrodef{iv}[IV]{Important Value}
\acrodef{sg}[SG]{Suspicious Graph}
\acrodef{mttd}[MTTD]{Mean Time to Detect}
\acrodef{soc}[SOC]{Security Operations Center}
\acrodef{nlp}[NLP]{Natural Language Processing}
\acrodef{apt}[APT]{Advanced Persistent Threat}
\acrodef{ioc}[IoC]{Indicator of Compromise}
\acrodef{gioc}[gIoC]{General Indicator of Compromise}
\acrodef{svo}[SVO]{Subject-Verb-Object}
\acrodef{ip}[IP]{Internet Protocol Address}
\acrodef{stp}[STP]{Steiner Tree Problem}
\acrodef{cti}[CTI]{Cyber Threat Intelligence}
\acrodef{edr}[EDR]{Endpoint Detection and Response}
\acrodef{stix}[STIX]{Structured Threat Information eXpression}
\acrodef{lotl}[LotL]{Living off the Land}

\acrodef{amid}[CKD]{CTI Knowledge Database}

\acrodef{aod}[AOD]{Attack Operation Descriptional}

\acrodef{attck}[ATT\&CK]{ATT\&CK}

\newcommand{\huawei}[1]{AnonymousCompany}

\acrodef{atie}[ATIE]{ATT\&CK Technique Information Entry}
\acrodefplural{atie}{ATT\&CK Technique Information Entries}

\newlength{\MaxSizeOfLineNumbers}%
\definecolor{keywordcolor}{rgb}{0.8,0.1,0.5}
\definecolor{lightlightgray}{gray}{.96}
\definecolor{lightgray}{gray}{.925}
\definecolor{medlightgray}{gray}{0.7}
\definecolor{medgray}{gray}{0.4}
\definecolor{darkgray}{gray}{0.35}
\definecolor{nearblack}{gray}{0.15}

\crefname{component}{Component}{Components}
\newcommand{\circleblacknum}[1]{%
  \tikz[baseline=(char.base)]{
    \node[shape=circle, fill=black, text=white, inner sep=1pt] (char) {\textbf{#1}};
  }%
}

\newtcblisting{promptbox}[2][]{%
    arc=2pt,
    outer arc=2pt,
    colback=gray!5!white,
    colframe=blue!50!black,
    fonttitle=\bfseries,
    title=#2,
    listing only,
    listing options={
        basicstyle=\small\ttfamily,
        keywordstyle={},
        stringstyle=\color{black},
        showstringspaces=false,
        numbers=none
    },
    left=5mm,
    right=5mm,
    top=3mm,
    bottom=3mm
}

\begin{document}
%
% paper title
% Titles are generally capitalized except for words such as a, an, and, as,
% at, but, by, for, in, nor, of, on, or, the, to and up, which are usually
% not capitalized unless they are the first or last word of the title.
% Linebreaks \\ can be used within to get better formatting as desired.
% Do not put math or special symbols in the title.
\title{KnowHow: Automatically Applying High-Level CTI Knowledge for Interpretable and Accurate Provenance Analysis}

% author names and affiliations
% use a multiple column layout for up to three different
% affiliations
%\author{\IEEEauthorblockN{Michael Shell}
%	\IEEEauthorblockA{Georgia Institute of Technology\\
%		someemail@somedomain.com}
%	\and
%	\IEEEauthorblockN{Homer Simpson}
%	\IEEEauthorblockA{Twentieth Century Fox\\
%		homer@thesimpsons.com}
%	\and
%	\IEEEauthorblockN{James Kirk\\ and Montgomery Scott}
%	\IEEEauthorblockA{Starfleet Academy\\
%		someemail@somedomain.com}}
	
% conference papers do not typically use \thanks and this command
% is locked out in conference mode. If really needed, such as for
% the acknowledgment of grants, issue a \IEEEoverridecommandlockouts
% after \documentclass

% for over three affiliations, or if they all won't fit within the width
% of the page, use this alternative format:
% 
%\author{\IEEEauthorblockN{Michael Shell\IEEEauthorrefmark{1},
%Homer Simpson\IEEEauthorrefmark{2},
%James Kirk\IEEEauthorrefmark{3}, 
%Montgomery Scott\IEEEauthorrefmark{3} and
%Eldon Tyrell\IEEEauthorrefmark{4}}
%\IEEEauthorblockA{\IEEEauthorrefmark{1}School of Electrical and Computer Engineering\\
%Georgia Institute of Technology,
%Atlanta, Georgia 30332--0250\\ Email: see http://www.michaelshell.org/contact.html}
%\IEEEauthorblockA{\IEEEauthorrefmark{2}Twentieth Century Fox, Springfield, USA\\
%Email: homer@thesimpsons.com}
%\IEEEauthorblockA{\IEEEauthorrefmark{3}Starfleet Academy, San Francisco, California 96678-2391\\
%Telephone: (800) 555--1212, Fax: (888) 555--1212}
%\IEEEauthorblockA{\IEEEauthorrefmark{4}Tyrell Inc., 123 Replicant Street, Los Angeles, California 90210--4321}}

\author{\IEEEauthorblockN{Yuhan Meng\IEEEauthorrefmark{2},
Shaofei Li\IEEEauthorrefmark{2}, 
Jiaping Gui\IEEEauthorrefmark{3}, 
Peng Jiang\IEEEauthorrefmark{4}, 
and Ding Li\thanks{* is the corresponding author.}\IEEEauthorrefmark{2}\IEEEauthorrefmark{1}}
\IEEEauthorblockA{\IEEEauthorrefmark{2}Key Laboratory of High-Confidence Software Technologies (MOE), School of Computer Science, Peking University}
\IEEEauthorblockA{\IEEEauthorrefmark{3}School of Computer Science, Shanghai Jiao Tong University, \IEEEauthorrefmark{4}Southeast University}
\IEEEauthorblockA{\IEEEauthorrefmark{2}\{mengyuhan, lishaofei, ding\_li\}@pku.edu.cn, \IEEEauthorrefmark{3}jgui@sjtu.edu.cn, \IEEEauthorrefmark{4}pengjiang@seu.edu.cn}}
%\IEEEauthorblockA{\IEEEauthorrefmark{4}pengjiang@seu.edu.cn}}

% use for special paper notices
%\IEEEspecialpapernotice{(Invited Paper)}

\eat{
\IEEEoverridecommandlockouts
\makeatletter\def\@IEEEpubidpullup{6.5\baselineskip}\makeatother
\IEEEpubid{\parbox{\columnwidth}{
		Network and Distributed System Security (NDSS) Symposium 2026\\
		23 - 27 February 2026 , San Diego, CA, USA\\
		ISBN 979-8-9919276-8-0\\
		https://dx.doi.org/10.14722/yyy.2026.[23$|$24]xxx\\
		www.ndss-symposium.org
}
\hspace{\columnsep}\makebox[\columnwidth]{}}

}

\IEEEoverridecommandlockouts
\makeatletter\def\@IEEEpubidpullup{6.5\baselineskip}\makeatother
\IEEEpubid{\parbox{\columnwidth}{
		Network and Distributed System Security (NDSS) Symposium 2026\\
		23 - 27 February 2026 , San Diego, CA, USA\\
		ISBN 979-8-9919276-8-0\\  
		https://dx.doi.org/10.14722/ndss.2026.[23$|$24]xxxx\\
		www.ndss-symposium.org
}
\hspace{\columnsep}\makebox[\columnwidth]{}}

% make the title area
\maketitle

% As a general rule, do not put math, special symbols or citations
% in the abstract
\begin{abstract}
    High-level natural language knowledge in \ac{cti} reports, such as the ATT\&CK framework, is beneficial to counter \ac{apt} attacks. However, how to automatically apply the high-level knowledge in \ac{cti} reports in realistic attack detection systems, such as provenance analysis systems, is still an open problem. The challenge stems from the semantic gap between the knowledge and the low-level security logs: while the knowledge in \ac{cti} reports is written in natural language, attack detection systems can only process low-level system events like file accesses or network IP manipulations. Manual approaches can be labor-intensive and error-prone.
        
    In this paper, we propose \toolname, a \ac{cti}-knowledge-driven online provenance analysis approach that can automatically apply high-level attack knowledge from \ac{cti} reports written in natural languages to detect low-level system events. The core of \toolname is a novel attack knowledge representation, \ac{gioc}, that represents the subject, object, and actions of attacks. By lifting system identifiers, such as file paths, in system events to natural language terms, \toolname can match system events to \acp{gioc} and further match them to techniques described in natural languages. Finally, based on the techniques matched to system events, \toolname reasons about the temporal logic of attack steps, detects potential \ac{apt} attacks in system events, and generates the human-readable report for each alert. Our evaluation shows that \toolname can accurately detect all 16 \ac{apt} campaigns in the open-source and industrial datasets, while existing approaches all introduce large numbers of false positives. Meanwhile, our evaluation also shows that  \toolname reduces at most 90\% of node-level false positives while having a higher node-level recall and is robust against several unknown attacks and mimicry attacks.

\end{abstract}

 \section{Introduction}

\acf{apt} attacks are serious attacks in the modern world. Recently, multiple \ac{apt} attacks have caused massive losses~\cite{ukraine,splunk}. To counter \ac{apt} attacks, researchers have proposed provenance analysis techniques that mine system events to detect \ac{apt} attacks~\cite{provtalk, depcomm, shadewatcher, hassan2019nodoze, atlas, airtag,10.1145/3427228.3427255,wang2020you}. However, most existing provenance analysis techniques are data-driven, which inevitably introduce many false positives and make the detection result hard to interpret~\cite{ depcomm, wang2020you,hassan2019nodoze, shaofeiempirical}. For instance, when an attacker executes a Word document containing a macro virus~\cite{msword},  data-driven approaches may mistakenly report other Word processes as malicious and cannot explain why each ``msword'' process is considered malicious, which can be confusing for a security analyst encountering such malware for the first time.  This makes it challenging for security personnel to process all alerts in a timely manner, potentially leaving some unaddressed, which ultimately may pose risks to the system~\cite{shaofeiempirical,hassan2019nodoze,10.1145/3571726}.

Besides data-driven approaches, researchers also propose  \acf{cti} knowledge driven approaches~\cite{holmes,aptshield,poirot,attackg,trec}. These approaches 
analyze \ac{cti} reports (e.g., MITRE ATT\&CK framework~\cite{attckweb}, APT organization reports~\cite{apt29campaign2, apt37fireeye} and research blogs~\cite{keydnapcase, apt41case}), extract knowledge from these documents, and map the extracted knowledge to system events to detect attacks. Existing \ac{cti}-knowledge-driven approaches fall into two categories: \ac{ioc} based approaches~\cite{extractor,threatraptor,attackg, ttpdrill,rhoades2014machine} and high-level knowledge-based approaches~\cite{aptshield,trec,ladder}. The former one extracts low-level \acp{ioc}, such as concrete attack-related file names and IP addresses, and detects whether system events contain the reported \acp{ioc}. The latter one leverages human experts to curate detection rules or log patterns to detect attacks in system events.

Unfortunately, existing \ac{cti}-knowledge-driven approaches are limited due to the lack of extensibility. \ac{ioc}-based approaches ignore general information in \ac{cti} documents but only focus on specific \acp{ioc}, which quickly become outdated~\cite{kure2022integrated} and cannot detect the varients of known attacks. High-level knowledge-based approaches rely on manual rules, which are expensive to build and imprecise. Holistically, both types of approaches can hardly be extended to keep up with the rapidly evolving attack techniques.

The key challenge of mapping the knowledge contained in high-level \ac{cti} reports to low-level provenance data is the semantic gap between natural language descriptions and system events. For instance, the \ac{cti} description ``Lazagne carefully scanned the browser's resource directory to extensively steal the credential files of users" should correlate with system events like ``la1 read ../firefox/resource..", but the lack of linguistic similarity makes direct matching difficult. \ac{ioc}-based approaches cannot handle this case since the description contains no \ac{ioc}. For high-level knowledge-based approaches, building a mapping rule requires significant efforts from advanced human experts.

 Our insight to make \ac{cti}-knowledge-driven approaches more extensible is that the mapping from high-level \ac{cti} knowledge can be largely automated by focusing on core elements in \ac{cti} reports. Specifically, we focus on three core elements in \ac{cti} reports: \textit{attack conductors}, \textit{attack actions}, and \textit{attack targets}.  These three core elements can be precisely mapped to key elements in system events: \textit{processes}, \textit{system calls}, and \textit{operation files}. Therefore, by leveraging minimal lifting rules, system events can be automatically mapped to the three core elements in \ac{cti} reports using NLP techniques.

Based on our insight, we propose \toolname, a \ac{cti}-knowledge-driven provenance analysis approach that automatically applies high-level attack knowledge from \ac{cti} reports to detect \ac{apt} attacks in low-level system events. The core of our approach is \ac{amid} that features a new intermediate representation, the \acf{gioc}, to represent the three core components in \ac{cti} reports. To effectively match \acp{gioc} and system events, we propose a novel approach that lifts the semantics of system events to the natural language level  (e.g., lifting ``firefox'' to ``browser''), thereby bridging the gap between them. Besides, we introduce a novel attack reasoning approach that automatically identifies \ac{apt} attacks in system events based on \ac{amid}. Our attack reasoning approach aligns attack techniques with attack tactics and APT Lifecycle stages, thereby reducing false positives by eliminating attack steps that do not adhere to the temporal logic of APT attacks~\cite{holmes,lifecyclecrowded,lifecycleibm,lifecycledell,lifecyclemandiant,lifecycleswiss}. 
Critically, \toolname leverages low-level alert provenance graphs and lifecycles derived from the attack reasoning model to generate comprehensive APT reports that summarize the attack in terms of high-level context, technical details, and
actionable guidance, enabling analyst-centric output for rapid triage and response.

We conduct a comprehensive evaluation of \toolname using the most recent and widely adopted open-source datasets, along with an industrial dataset. \toolname precisely identifies all attack campaigns within our datasets, whereas all existing baselines exhibit much more false positives.  Specifically, \toolname reduces false positives at the node level by 81\% and 90\% for the two recent baselines, \nodlink~\cite{nodlink} and \kairos~\cite{kairos}, respectively. Regarding the effect of the extraction on \ac{apt} detection, our evaluation confirms that knowledge extracted using \toolname can help reduce false positives in downstream \ac{apt} attack detection tasks by 68\% and 79\%, respectively, compared to utilizing knowledge from \ladder and \extractor—two of the latest \ac{cti} information extraction frameworks. Our evaluation further shows that \toolname's novel \ac{cti} knowledge condensation representation, \ac{gioc}, is the primary reason for its superiority over existing approaches. Moreover, it has been deployed in open-world scenarios, with the experimental results collectively validating the practical usability of \toolname.

We summarize our contributions as follows:
\begin{itemize}[leftmargin=*]
    \item We propose \toolname, a \ac{cti}-knowledge-driven online provenance analysis solution that can automatically apply high-level attack knowledge from \ac{cti} reports to detect \ac{apt} attacks in low-level system events.

    \item We introduce a novel compact representation, \ac{gioc}, for high-level attack knowledge and manage it within a novel knowledge base, \ac{amid}.
    \item We design a novel attack reasoning method based on the stages of the APT Lifecycle identified by querying \ac{amid}.
    \item We thoroughly evaluate \toolname on widely used datasets, as well as an industrial dataset. The results confirm that \toolname meets the requirements for accurate, efficient, and interpretable detection of APT attacks.
    \item We have deployed \toolname within the OpenEuler ecosystem, which is a production-grade, community-driven Linux distribution, to verify its validity in practical environments. 
    \item In response to the practical need for further improved alert interpretability, as expressed by security practitioners in industry, we have extended \toolname's workflow with an automated APT report generation module powered by LLMs in this extended version. This extension further analyzes and systematizes the generated alerts, enhancing their comprehensibility and operational utility.
\end{itemize}

\section{Threat Model and Assumptions}
We follow the same threat model used in previous works on system monitoring~\cite{depimpact, depcomm, shadewatcher, poirot, holmes, provtalk}.
Specifically, we assume the kernel and kernel-layer auditing frameworks~\cite{sysdig, etw, linuxaduit} are not compromised. Any kernel-level attack campaign is outside the scope. We assume that there is an external attacker, who attacks the victim system remotely. Based on the above assumptions, the attacker can only achieve system intrusion by inducing the victim to download and execute a malicious payload or by exploiting a vulnerability.

\section{Background and Motivation}

 \begin{figure*}[t!]
     \centering
     \includegraphics[width=\linewidth]{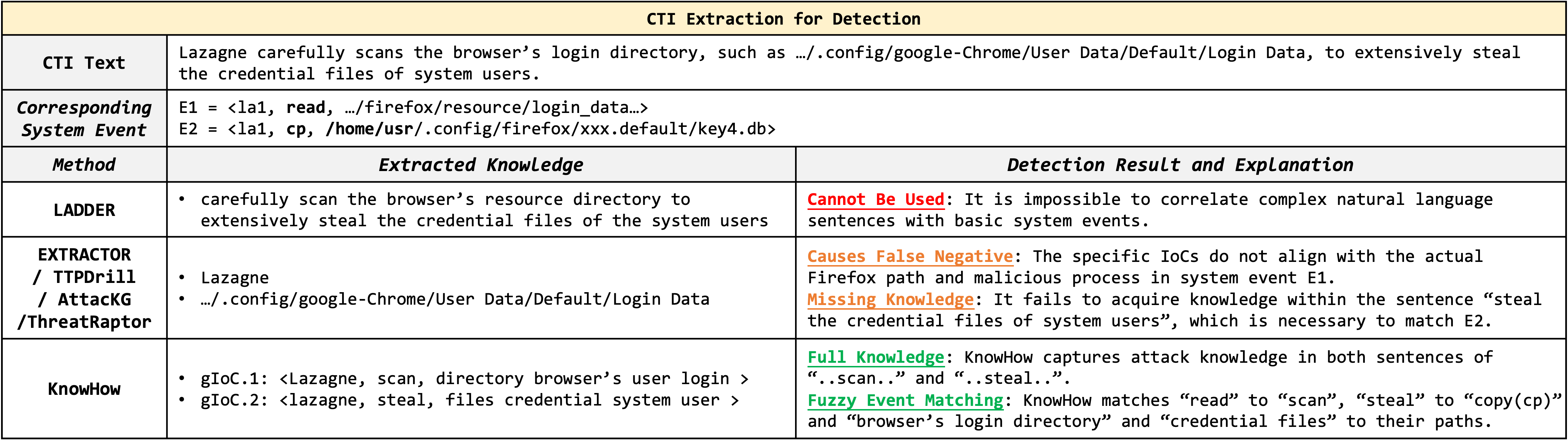}
     \caption{A comparison of various \ac{cti}-knowledge driven methods with the detection results using the extracted knowledge on the Lazagne case\cite{lazagne-git,lazagne2}. }
 	\label{fig:insight}
 \end{figure*}

A \ac{cti}-knowledge-driven approach has two stages~\cite{extractor, threatraptor,ttpdrill,stix,ladder, aptshield}: (1) extracting knowledge from \ac{cti} reports and (2) mapping the extracted knowledge to low-level system events. In practice, the extracted knowledge may also be organized in standard formats, such as \ac{stix}~\cite{stix}. Based on the types of extracted knowledge, we categorize existing \ac{cti}-knowledge-driven approaches into \ac{ioc}-based approaches and high-level knowledge-based approaches. We show the differences between these two types of \ac{cti} knowledge in Figure~\ref{fig:insight}.

\subsection{\ac{ioc}-Based Approaches}
\ac{ioc}-based approaches, such as \extractor~\cite{extractor}, \ttpdrill~\cite{ttpdrill}, and \threatraptor~\cite{threatraptor}, extract concrete \acp{ioc} from \ac{cti} reports and match them against entities observed in system events. For instance, given the \ac{cti} text shown in Figure~\ref{fig:insight}, these methods extract specific \acp{ioc} such as the malicious process name \textit{Lazagne} and the suspicious file path \textit{.../.config/google-chrome/User Data/Default/Login Data}. These \acp{ioc} are then used to match entities in system events for attack detection.

\ac{ioc}-based approaches suffer from high false-negative rates due to their reliance on static, time-sensitive indicators that quickly become obsolete and fail to generalize across environments (e.g., ``Lazagne" vs ``la1" in Figure~\ref{fig:insight}), while also exhibiting knowledge omission by ignoring valuable behavioral descriptions in favor of concrete artifacts. 
For example, \ac{ioc} only includes the Chrome browser path as Lazagne’s target, ignoring the description “scan the browser's login directory,” thus failing to detect attacks on other browsers like Firefox in Figure~\ref{fig:insight}.

\subsection{High-Level Knowledge-Based Approaches}
High-level knowledge-based approaches recognize that \ac{cti} reports often contain rich, descriptive content about attack behaviors and patterns beyond simple \acp{ioc}. These methods aim to extract such high-level knowledge from natural language CTI reports~\cite{ladder} and convert it into actionable detection rules through expert interpretation~\cite{aptshield, holmes}. An example of extracted high-level knowledge from one of these approaches, \ladder, is shown in Figure~\ref{fig:insight}.

The key challenge is how to convert high-level knowledge into actionable rules. To this end, researchers have proposed many interesting methods that rely on expert-curated rules. Specifically, \holmes~\cite{holmes} and \aptshield~\cite{aptshield} perform pattern matching by checking whether observed system behaviors conform to their predefined attack lifecycle rules and sensitivity labels, while \rapsheet~\cite{rapsheet} uses the manually curated rules to flag suspicious activities based on known attack indicators. 
\poirot~\cite{poirot} and \attackg~\cite{attackg} use expert-curated query graphs and attack technique templates derived from \ac{cti} reports, which are then applied to system events for detection.
Likewise, \trec~\cite{trec} performs graph-level feature matching between selected procedures from the Atomic Red Team~\cite{secbudget} and system events to detect potential threats.

The limitation of existing high-level knowledge-based approaches is that they rely on static, concrete rules, which are often too strictly defined and specific to individual systems. This limitation hinders their ability to generalize to new systems and adapt to evolving attack patterns. For example, \holmes, \rapsheet, and \aptshield rely on manually defined whitelists, such as trusted IP addresses, benign command lists, sensitive files, and critical system files. These whitelists are often tailored to specific environments and require continuous updates to remain effective. Moreover, \poirot, \attackg, and \trec focus on concrete attack procedures extracted from \ac{cti} reports, which are often instance-specific and lack sufficient abstraction, limiting their ability to handle diverse and evolving attack variants.

\subsection{\acl{stix}}
\ac{stix} provides a standardized way to represent both concrete \acp{ioc} and high-level attack descriptions (e.g., TTPs, Attack Patterns). \ac{stix}, by standardizing the representation of \ac{cti} knowledge, facilitates knowledge sharing across organizations. However, how to effectively use the full knowledge in \ac{stix} is still an open problem. 

Moreover, although \ac{stix} supports representing high-level knowledge in its design, the community uaually uses \ac{stix} to share \acp{ioc} instead of high-level attack knowledge. A study shows that over 94.93\% of \ac{stix} objects contain only \ac{ioc} attributes, while less than 0.1\% include high-level behavioral descriptions~\cite{cti-lense}.

\subsection{Motivation and Insight}

Our key insight for an extensible high-level knowledge-based approach is that subject-verb-object phrases in CTI reports encapsulate richer attack knowledge that can be structured into more meaningful and extensible representations.
These phrases decompose into three fundamental components: \textit{attack conductors} (who), \textit{attack actions} (what operation), and \textit{attack targets} (on what), reflecting the core semantic structure of attack behaviors, exemplified by ``la1 reads login data files'' in Figure~\ref{fig:insight}.

Unlike static and context-free \acp{ioc}, this structured representation preserves richer semantic information by including not only the core entities but also their relationships and contextual descriptions. As shown in Figure~\ref{fig:insight}, \textit{``steal the credential files of system user''} contains both the conductor (\textit{lazagne}), the action (\textit{steal}), and the target (\textit{credential files of system user}), which provides a more comprehensive understanding of the attack behavior. 
The extracted knowledge can then be used to match with the system events \texttt{<la1, read, .../firefox/resource/login\_data...>} in a proper way to detect the attack, which can address the limitation of existing high-level knowledge extraction methods. Compared to \ac{stix}, our representation contains more structured and actionable contextual information, which is essential for understanding the attack behaviors.
Therefore, it can be used as a general structured knowledge representation that bridges the gap between high-level \ac{cti} knowledge and low-level system events, enabling more effective and automated attack detection.

\section{CTI Knowledge Database}
\label{sec:amid}

To bridge the semantic gap between high-level knowledge in \ac{cti} reports and low-level system events, we propose \ac{amid}, a knowledge database dedicated to high-level attack knowledge derived from \ac{cti} reports.
\ac{amid} is defined as a set of \acp{atie}, each representing an attack technique as defined by MITRE ATT\&CK~\cite{attckweb}. An \ac{atie} comprises four fields: a unique ID for a technique in ATT\&CK ($uid$), a description of the technique ($des$), a \ac{cti} list ($list_{cti}$),  and a \ac{gioc} list ($list_{gioc}$). Specifically, $list_{cti}$ contains \ac{cti} reports in which the technique can be identified. $list_{gioc}$ contains \acp{gioc}, which encapsulate high-level attack knowledge reported in the corresponding \ac{cti} reports in $list_{cti}$ and will be elaborated on in the following subsections.
To support high performance \ac{apt} detection, \ac{amid} also provides a provenance query, $ProvQ$, that enables users to identify \acp{atie} related to a given system event.

\subsection{General Indicators of Compromises}
\label{sec:gioc}
\ac{gioc} is the key component in \ac{amid} that connects high-level attack knowledge in \ac{cti} reports with low-level system events. Conceptually, a \ac{gioc} is a compact representation of attack descriptive sentences in \ac{cti} reports and can be automatically learned from \ac{cti} reports using \ac{nlp} techniques. Formally, a \ac{gioc} is a \ac{svo} triplet structure, \textit{(subject, verb, object)}, which captures the core information of "who performs which operation on what" in attack descriptions. A typical \ac{gioc} is shown in Figure \ref{fig:insight}.

Specifically, the \textit{subject} is the conductor of an attack. It can be the name of an attacker, malware, or a hijacked application that initiates attacks. For example, in the \ac{cti} sentence, ``APT41 used built-in commands \textit{net} to enumerate local administrator groups''\cite{apt41case}, ``APT41'' is the attacker subject. In contrast, in another \ac{cti} sentence, ``Keydnap adds the \textit{setuid} flag to a binary to easily elevate in the future''\cite{keydnapcase}, ``Keydnap'' is the malware subject. The $verb$ represents the action of attacks, such as ``use'' and ``enumerate'' in the APT41 example, and ``add'' and ``elevate'' in the Keydnap example. Finally, \textit{object} is the target of an attack, such as ``built-in commands \textit{net}'' and ``local administrator groups'' in the APT41 example, and ``the \textit{setuid} flag'' and ``a binary'' in the Keydnap example.

The key difference between \acp{gioc} and conventional SVO pairs is that the \textit{subjects} and \textit{objects} of \acp{gioc} are attack-relevant concepts and their associated information, such as modifiers describing these concepts. On a high level, we define a noun as an attack-relevant concept if it constitutes an \ac{ioc} or the name of a system object (e.g., an application name, domain names, file names, etc.). Our emphasis on attack-relevant concepts stems from our objective to correlate high-level \ac{cti} reports with low-level system events. These concepts provide clues about tools, IP addresses, and files that attackers might utilize during an attack, which can facilitate the matching of low-level events. By recognizing attack-relevant concepts, we can avoid overly broad statements like ``the attack originated from state-owned groups,'' which are less useful for automated low-level event matching. Note that attack-relevant nouns are more versatile than conventional \acp{ioc}, allowing our approach to match low-level events with corresponding \ac{gioc} even when the textual content does not match precisely.

Formally, we define a noun $N$ as an attack-relevant concept if it fulfills one of the following five conditions:
\begin{enumerate}[label=\arabic*), leftmargin=*]
\item $N$ is an \ac{ioc}, such as a file name, an IP address, a file hash, etc.
\item $N$ is a domain name.
\item $N$ is the name of an application or malware.
\item $N$ is a command (e.g., cp) or its full name (e.g., copy)
\item $N$ represents a general concept of system objects, including but not limited to terms like ``file,'' ``directory,'' ``IP address,'' ``process,'' ``application,'' ``registry,'' and their synonyms.
\end{enumerate}

We propose conditions 2-5 to enhance the extensibility of \acp{gioc}. For example, condition 5 allows our approach to capture expressions like ``browser's folder'' in Figure~\ref{fig:insight}, thereby facilitating the matching of events involving Firefox with attack knowledge pertinent to Chrome.

Given an attack-relevant concept, we also consider modifiers (e.g.,  ``browser data'') and subclauses (e.g., ``steal the credential files of system users'') that describe the concept and serve as its related information. This is due to the fact that nouns in \ac{cti} reports are often devoid of context and meaning. For example, in Figure~\ref{fig:insight}, the identified attack-relevant concept in the sentence ``..scans the browser’s resource directory.. '' is ``directory'', which carries limited significance. Instead, we must identify the related information, such as ``browser data'' and ``steal the credential files of system users'' to align with low-level events involving access to Firefox's user data and system credential files.

Compared with \acp{ioc}, the key advantage of \acp{gioc} lies in their ability to reflect higher-level information. \acp{gioc} extract attack information from a behavioral perspective, whereas \acp{ioc} focus on an instance perspective. Therefore, \acp{gioc} can be better extended across different environments. For example, a typical type of \ac{ioc} is the process name of malware appearing in system events. Consequently, \ac{ioc} knowledge can only identify malware with the same process name in the event, ignoring other pertinent information. In contrast, \acp{gioc} summarize attack behavioral information and match it across various parts of the event, such as the target file of the malware or executed commands. Hence, they are more effective in detecting different variants of the malware.

\subsection{Building \ac{amid} from \ac{cti} Reports}
\label{build}

We leverage \ac{nlp} techniques to automatically identify concepts relevant to attacks. For \acp{ioc}, we use the IoCParser~\cite{iocparser} to extract \acp{ioc} more effectively. However, we observe that the original IoCParser lacks adequate support for identifying file paths, registry run keys, and command line operations. Therefore, \toolname extends the regular expressions in IoCParser to improve the identification of file paths and registry run keys. Meanwhile, \toolname maintains a list of commonly used command lines to recognize command line entries in \ac{cti}.  Table~\ref{tab:ioc_parse} illustrates the regular expressions and command line samples that we have extended. For conditions 2 and 3 mentioned in Section~\ref{sec:gioc}, we rely on large language models to identify domain and application names. Specifically, given a noun, we submit it to a large language model and inquire whether it is a domain name or an application/malware name. For condition 4, we maintain a list of common Linux and Windows system commands, along with their corresponding full names, based on their documentation. For condition 5, we maintain a list of general concepts of system objects and detect their synonyms using large language models. In our approach, we can leverage existing NLP toolkits, such as Stanford CoreNLP~\cite{stanfordnlp}, to identify the modifiers and subclauses associated with an attack-relevant concept.

\begin{table}[]
\centering
\caption{Excerpts of the Extended \ac{ioc} Parsing Rules.}
\label{tab:ioc_parse}
% \resizebox{0.98\columnwidth}{!}{

\label{tab:robust}
\begin{adjustbox}{max width=\linewidth}
\renewcommand{\arraystretch}{1.0} 
%\normalsize
\begin{tabular}{c|c|c} \hline
                                                                                & \textbf{Example}                                         & \textbf{Regular Expression}       \\ \hline
\multirow{9}{*}{\begin{tabular}[c]{@{}c@{}}Registry\\Run\\Key\end{tabular}} 
                                                                               & \multirow{9}{*}{\begin{tabular}[c]{@{}c@{}}HKCU\textbackslash{}\\Software\textbackslash{} \\ Microsoft\textbackslash{}\\Windows NT\textbackslash{} \\ CurrentVersion\textbackslash{}\\Winlogon \end{tabular}} &
                                                                               \multirow{9}{*}{\begin{tabular}[c]{@{}c@{}}\textasciicircum{}(HKCU\textbar HKLM\textbar HKCR\textbar\\ HKCC\textbar 
                                                                               HKEY\_LOCAL\_\\MACHINE\textbar 
                                                                               HKEY\_\\CURRENT\_USER\textbar \\
                                                                               HKEY\_CLASSES\\\_ROOT\textbar 
                                                                               HKEY\_\\CURRENT\_CONFIG\textbar \\
                                                                               HKEY\_USERS) \\
                                                                               (\textbackslash{}\textbackslash{}{[}\textasciicircum{}\textbackslash{}\textbackslash{}?/*\textbar \textless{}\textgreater{}:"{]}+)*\$\end{tabular}} \\ 
                                                                               &                                                  &                       \\
                                                                               &                                                  &                       \\
                                                                               &                                                  &                       \\
                                                                               &                                                  &                       \\
                                                                               &                                                  &                       \\
                                                                               &                                                  &                       \\
                                                                               &                                                  &                       \\
                                                                               &                                                  &                       \\ \hline
\multirow{7}{*}{\begin{tabular}[c]{@{}c@{}}File\\Name,\\File\\Path\end{tabular}} 
                                                                               & \multirow{7}{*}{\begin{tabular}[c]{@{}c@{}}C:\textbackslash{}\textbackslash{}Windows\textbackslash{}\\Microsoft.NET  \textbackslash{}\\Framework64\textbackslash{}, \\ /root/.vscode\\-server/ \end{tabular}}  &
                                                                                \multirow{7}{*}{\begin{tabular}[c]{@{}c@{}} \textasciicircum{}(?:(?:{[}a-zA-Z{]}:\textbar \textbackslash{}.\\ \{1,2\})? 
                                                                                {[}\textbackslash{}\textbackslash{}/{]}(?:{[}\textasciicircum{}\textbackslash{}\textbackslash{}\\?/*\textbar \textless{}\textgreater{}:"{]}+ {[}\textbackslash{}\textbackslash{}\\/{]})*)(?:(?:{[}\textasciicircum{}\textbackslash{}\textbackslash{}?/*\\ \textbar \textless{}\textgreater{}:"{]}+?) \\ 
                                                                                (?:\textbackslash{}.{[}\textasciicircum{}.\textbackslash{}\textbackslash{}?/*\textbar \\ \textless{}\textgreater{}:"{]}+)?)?\$\end{tabular}}  \\ 
                                                                               &                                                  &                       \\
                                                                               &                                                  &                       \\
                                                                               &                                                  &                       \\
                                                                               &                                                  &                       \\
                                                                               &                                                  &                       \\
                                                                               &                                                  &                       \\ \hline
\multirow{7}{*}{\begin{tabular}[c]{@{}c@{}}Typical\\Linux\\Command \end{tabular}} 
                                                                                & \multirow{7}{*}{\begin{tabular}[c]{@{}c@{}}scp, ssh, sftp, tftp,\\ curl, sshd,
                                                                                certutil,\\ wget, ls, rm, sh,\\ mv,
                                                                                stat, cat,\\ reg add, reg del,\eat{, arp}\\
                                                                                kill, pkill, grep,\\ find, cat, ifconfig\\                                                               
                                                                                \end{tabular}}  & \multirow{7}{*}{ Same as IoCParser } \\
                                                                                & & \\
                                                                                & & \\
                                                                                & & \\
                                                                                & & \\
                                                                                & & \\
                                                                                & & \\ \hline
\multirow{7}{*}{\begin{tabular}[c]{@{}c@{}}Typical\\Windows\\Command\end{tabular}} 
                                                                                & \multirow{7}{*}{\begin{tabular}[c]{@{}c@{}}Get-Process,\\ Get-Service,\\ Get-ChildItem,\\
                                                                                New-Item,\\ Remove-Item,\\ Set-Location,\\
                                                                                Clear-Host\\ 
                                                                                \eat{Write-Host\\Get-EventLog, 
                                                                                Stop-Process\\ Start-Service, Restart Service\\
                                                                                ipconfig, del, schtask\\ tasklist, dir\\}
                                                                                \end{tabular}}  & \multirow{8}{*}{ Same as IoCParser } \\ 
                                                                                & & \\
                                                                                & & \\
                                                                                & & \\
                                                                                & & \\
                                                                                & & \\
                                                                                
                                                                                & & \\ \hline

\end{tabular}
%}
\end{adjustbox}
\end{table}

\noindent\textbf{\acp{gioc} Extraction.} After identifying and complementing the attack-relevant concepts, our approach extracts \acp{gioc} from \ac{cti} reports. It first extracts sentences that contain these concepts identified in the first step. Then, for each of these sentences, we employ SVO toolkits to extract \textit{(subject, verb, object)} tuples, which serve as the backbones of \acp{gioc}. For example, for the sentence shown in Figure~\ref{fig:insight}, the extracted backbone is \textit{(Lazagne, scan, directory)}.

Then, we enhance the semantics of the extracted backbone by appending the modifier text to its corresponding nouns. For example, as shown in Figure~\ref{fig:insight}, the enriched \ac{gioc} would be \textit{(Lazagne, scan, directory browser's user login)}. Here, we do not take into account the order or grammatical correctness of the text since these elements are adequate for modern NLP techniques to generate sentence embeddings that can fuzzily match low-level events, as will be discussed in Section~\ref{sec:query}.
Specifically, we treat subclauses as regular sentences and process them in the same way as mentioned above. Therefore, the subclauses will also generate \acp{gioc}. 

There are two challenges in generating \ac{gioc}. 
The first challenge is the potential absence of certain components of \textit{(subject, verb, object)} in the subclauses. For example, in the sentence depicted in Figure~\ref{fig:insight}, the subclause ``to steal the credential files
of system users'' lacks a subject. To address this, \toolname directly borrows the missing component from the main clause to form complete \textit{(subject, verb, object)} tuples. 
The second challenge is that many attack-relevant concepts are specific \acp{ioc}, such as concrete file names and IP addresses, which cannot be extended for broad attack detection. To address this issue,  our approach adopts a rule-based method to extend common patterns of \acp{ioc} into \acp{gioc}. Our high-level strategy involves extracting components from \acp{ioc} that carry semantic meaning. For example, for the file name ``\textit{/usr/bin/long\_file\_name\-/wallet.db}'', our approach converts it into a \ac{gioc} by using the first-level folder name as ``the user folder'' and its file name as ``wallet db file''. This conversion allows a more extensible matching of files with similar names within BOTH Linux and Windows user folders. The excerpt of the conversion rules is outlined in Table~\ref{tab:simple-table}.

\begin{table}[h]
\centering
\caption{Excerpts of the simplification table to turn \acp{ioc} into \acp{gioc} , where $D$ means the name of this level dictionary, $F$ means the name of the files, $E$ means the extension name of the file, and $Dom$ means the DNS domain name of the corresponding IP. The simplification of directories is similar to that of files, except that there is no extension name.}
\label{tab:simple-table}
%\resizebox{0.98\columnwidth}{!}{%
\begin{adjustbox}{max width=\linewidth}
\huge
\begin{tabular}{c|c|c} \hline
                                                                               & \textbf{System Identifier}                                         & \textbf{Lifted Sentences}            \\ \hline
\multirow{17}{*}{\begin{tabular}[c]{@{}c@{}}Linux \\ File\end{tabular}}                                                        & /etc/$D$/*/$F$.$E$                               & etc $D$ $E$ file       \\ \cline{2-3} 
                                                                               & /var/$D$/*/$F$.$E$                               & var $D$ $E$ file       \\ \cline{2-3} 
                                                                               & /proc/[PID]/$D$/*/$F$.$E$                               & proc $D$ $E$ file       \\ \cline{2-3} 
 &
  \multirow{6}{*}{\begin{tabular}[c]{@{}c@{}}/bin/$D$/*/$F$.$E$,\\ /sbin/$D$/*/$F$.$E$,\\ /usr/bin/$D$/*/$F$.$E$, \\ /usr/sbin/$D$/*/$F$.$E$, \\ /usr/local/bin/$D$/*/$F$.$E$, \\ /usr/local/sbin/$D$/*/$F$.$E$\end{tabular}} &
  \multirow{6}{*}{$F$ $E$ file} \\
                                                                               &                                                  &                        \\
                                                                               &                                                  &                        \\
                                                                               &                                                  &                        \\
                                                                               &                                                  &                        \\ 
                                                                               
                                                                          &                                                  &                        \\       \cline{2-3} 
                                                                               & /home/aa/$D/*/$F$.$E\$                           & user $D$ $F$ $E$ file      \\ \cline{2-3} 
                                                                               & /root/$D$/*/$F$.$E$                              & root user $D$ $F$ $E$ file \\ \cline{2-3} 
 &
  \multirow{5}{*}{\begin{tabular}[c]{@{}c@{}}/lib/$D$/*/$F$.$E$,\\ /lib32/$D$/*/$F$.$E$,\\ /lib64/$D$/*/$F$.$E$, \\ /usr/local/lib/$D$/*/$F$.$E$, \\ /xx/lib/$D$/*/$F$.$E$\end{tabular}} &
  \multirow{5}{*}{$D$ library file} \\
                                                                               &                                                  &                        \\
                                                                               &                                                  &                        \\
                                                                               &                                                  &                        \\ 
                                                                                &                                                  &                        \\ \cline{2-3} 
                                                                               & other: */$F$.$E$                                 & $E$ file               \\ \hline
\multirow{8}{*}{\begin{tabular}[c]{@{}c@{}}Windows \\ File\end{tabular}} &
  \multirow{3}{*}{\begin{tabular}[c]{@{}c@{}}HKEY\_*, HKCU*, \\ HKCR\textbackslash{}*, HKLM*, \\ HKU*, HKCC*\end{tabular}} &
  \multirow{3}{*}{registry run key} \\
                                                                               &                                                  &                        \\
                                                                               &                                                  &                        \\ \cline{2-3} 
 &
  c:\textbackslash{}\textbackslash{}windows\textbackslash{}system32\textbackslash{}$D$\textbackslash{}*\textbackslash{}$F$.$E$ &
  windows system $D$ $F$.$E$ file \\ \cline{2-3} 
 &
  c:\textbackslash{}\textbackslash{}windows\textbackslash{}$D$\textbackslash{}*\textbackslash{}$F$.$E$ &
  windows system $D$ $F$.$E$ file \\ \cline{2-3} 
 &
  \multirow{2}{*}{\begin{tabular}[c]{@{}c@{}}c:\textbackslash{}\textbackslash{}ProgramFiles\textbackslash{}$D$\textbackslash{}*\textbackslash{}$F$.$E$, \\ c:\textbackslash{}\textbackslash{}ProgramFiles(x86))\textbackslash{}$D$\textbackslash{}*\textbackslash{}$F$.$E$\end{tabular}} &
  \multirow{2}{*}{$D$ $F$ $E$ file} \\
                                                                               &                                                  &                        \\ \cline{2-3} 
                                                                               & other: *\textbackslash{}$F$.$E$ & $F$ $E$ file               \\ \hline
\multirow{2}{*}{IP} &
  \begin{tabular}[c]{@{}c@{}}10.0.0.0/8, 172.16.0.0/12,\\ 192.168.0.0/16\end{tabular} &
  internal network \\ \cline{2-3} 
                                                                               & other                                & external network  $Dom$     \\ 
                                                                         \hline
\multirow{19}{*}{\begin{tabular}[c]{@{}c@{}}Command \\ Operation\end{tabular}} & cp                                               & copy                   \\ \cline{2-3} 
                                                                               & 
                                                                         \begin{tabular}[c]{@{}c@{}}    scp, ssh, sftp, tftp,\\ curl, sshd, certutil\end{tabular} &
  transfer \\ \cline{2-3}  
                                                                               & wget                                             & download               \\ \cline{2-3} 
                                                                               & ls,dir                                           & list                   \\ \cline{2-3} 
                                                                               & rm,del,rmdir                                     & remove                 \\ \cline{2-3} 
                                                                               & sh                                               & shell                  \\ \cline{2-3} 
                                                                               & stat, cat                                         & show                   \\ \cline{2-3} 
                                                                               & schtask                                          & schdule                \\ \cline{2-3} 
                                                                               & rundll                                           & run, dll file          \\ \cline{2-3} 
                                                                               & reg add                                          & add                    \\ \cline{2-3} 
                                                                               & reg del                                          & del                    \\ \cline{2-3} 
                                                                               & kill, pkill, taskkill                              & stop                   \\ \cline{2-3} 
                                                                               & grep, find                                        & search                 \\ \cline{2-3} 
                                                                               & cat                                              & read                   \\ \cline{2-3}
                                                                               & \multirow{4}{*}{\begin{tabular}[c]{@{}c@{}}powershell command with \\capital and lower-case letter \\ (\eg Invoke-Command, \\ Get-ChildItem)\end{tabular}}                         & \multirow{4}{*}{\begin{tabular}[c]{@{}c@{}}Divided by ``-'' and \\ capital letters. \\ (\eg Invoke, Command; \\Get, Child Item)\end{tabular}} \\
                                                                               &                                                  &                        \\
                                                                               &                                                  &                        \\
                                                                               &                                                  &                        \\ \hline
\multirow{4}{*}{\begin{tabular}[c]{@{}c@{}}System \\ Call\end{tabular}}        & execve                                           & execute                \\ \cline{2-3} 
                                                                               & recvmsg, recvfrom                                 & receive                \\ \cline{2-3} 
                                                                               & sendmsg, sendto                                   & send                   \\ \cline{2-3} 
                                                                               & chmod                                            & change, file mode      \\ \hline
\end{tabular}%
%}
\end{adjustbox}
\end{table}

\noindent\textbf{\acp{atie} Generation.} The final step of our approach is converting the extracted \acp{gioc} to \acp{atie}. This step needs first adding the ATT\&CK technique IDs and descriptions to the corresponding fields within the \ac{atie}.

For the ATT\&CK technique IDs and descriptions, our approach first scans to determine whether the \ac{cti} report references MITRE ATT\&CK technique descriptions~\cite{attckweb}. Specifically, the content is extracted using the PyMuPDF~\cite{PyMuPDF}, and regular expressions are employed to scan the technique IDs and descriptions. Note that numerous popular threat intelligence sources include ATT\&CK technique labels, which are explicitly summarized in their \ac{cti} reports~\cite{apt41all, hijack1, svchost1}. If the report explicitly cites a MITRE ATT\&CK technique, our approach generates an \ac{atie} for that technique.

If a \ac{cti} report does not explicitly reference an ATT\&CK Technique, our approach attempts to associate it with an existing \ac{atie}. To accomplish this, we have developed a function akin to $ProvQ$, which will be elaborated on in Section~\ref{sec:query}, to query the \ac{amid} for the most similar list of \acp{atie}. The sole difference is that, rather than utilizing a system event, we query the \ac{amid} using \acp{gioc} derived from the \ac{cti} reports. We also employ a similar method to compute similarity scores, as outlined in Section~\ref{sec:query}.

\subsection{Querying \ac{amid} with System Events}
\label{sec:query}
\ac{amid} provides the provenance query, $ProvQ$, that matches the low-level system events to \acp{atie}. The key advantage of $ProvQ$ is that it allows fuzzy matching to \acp{gioc}, providing a more general detection even when \acp{ioc} fail to match system events.

$ProvQ$ takes a system event $e=(source, destination, syscalltype$, $commandline)$ and returns a list of \acp{atie} that match the given event. We say that $e$ matches an \ac{atie}, $t$, if their similarity score is greater than the given query threshold ${\theta}_q$. The similarity score $Sim(e, t)$ between $e$ and $t$ is defined as $score(e, t) = \sum S(e.y, t)$, where $y$ is one of ``source'', ``destination'', ``syscalltype'', and ``commandline'', and $S(e.y, t)$ is the occurrences of \acp{gioc} in $t$ that appear in $e.y$, respectively.

\begin{figure}[t]
    \centering
    \includegraphics[width=\linewidth]{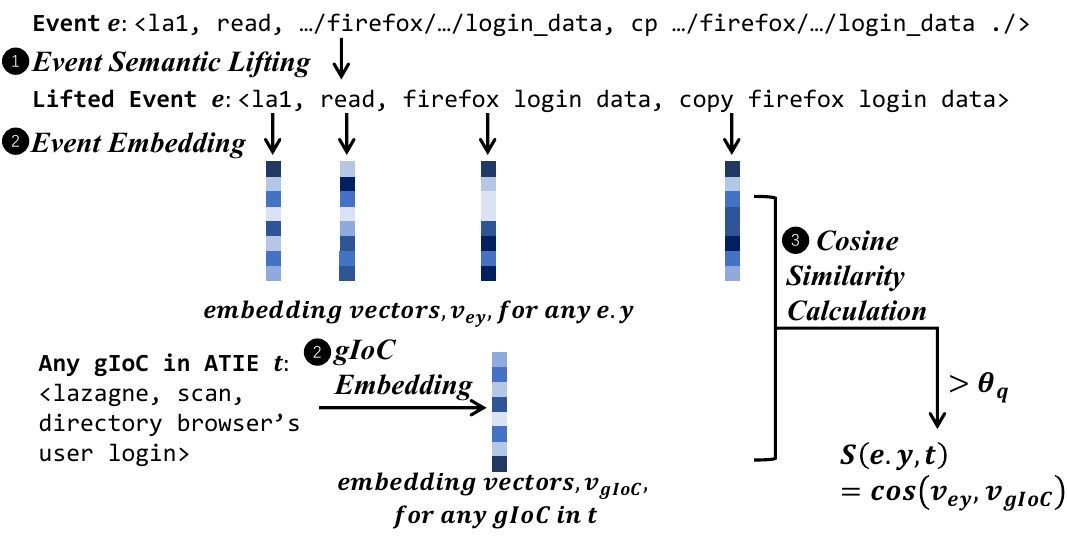}
	\caption{The calculation process of $S(e.y, t)$.}
	\label{fig:computing}
\end{figure}

 We calculate $S(e.y, t)$ by matching the semantics of $e.y$ to the subject-verb-object triplet of the \ac{gioc}. The high-level process is shown in Figure~\ref{fig:computing}.
The challenge is how to determine the occurrence of $t$ in $e.y$ since the natural language words in \ac{gioc} cannot directly match the system identifiers, such as file paths, in system events. 
To address this problem, \toolname first lifts the specific system events to extensible semantic representations and then uses semantic aware embedding to convert the system events and \acp{gioc} into numeric vectors. Finally, our approach calculates the cosine similarity of the embedding vectors of the system events and \acp{gioc}. If the cosine similarity is above a threshold, our approach considers the event ``hits'' a \ac{gioc}.

\noindent\textbf{Event Semantic Lifting.} 
The key step in constructing $ProvQ(e, {\theta}_q)$ involves elevating low-level system events to natural language sentences that can be matched with \acp{gioc}.  The high-level idea is the same as the conversion method mentioned in \ac{gioc} extraction in Section~\ref{build}. For file paths, we extract the file type name, application name, and the necessary path information as the semantic representation. The file type name can be inferred from the file's extension, while the application name is derived from its installation location within specific folders (e.g. the name of the folder directly under ``/bin'' is typically the application name). For IP addresses, we utilize their domain name (when available) in reverse DNS as the semantic representation. For example, ``64.233.160.0'' is represented as ``Google''. For IPs whose DNS domain cannot be resolved, we label them as ``unknown network''. Finally, for command lines and system call types, we expand abbreviations to their full forms for semantic representation. For example, we expand the command ``cp'' to ``copy'' and the system call ``execve'' to ``execute''.

\noindent\textbf{Event and \ac{gioc} Embedding.} After lifting the system events into natural language representations, \toolname converts system events and \acp{gioc} into numerical vectors so that we can calculate their cosine similarity. To this end, we leverage FastText~\cite{fasttext} to embed the lifted event entries and fields in \acp{gioc} into numerical vectors. We chose FastText due to its ability to preserve semantics and be efficient. First, embedding vectors of FastText can preserve the semantics of words. For example, FastText ensures the distances between ``run'' and ``execute'' to be small so that we can match the system call ``execute'' to text ``run'' in \ac{cti} reports. Second, compared to other techniques like large language models, FastText is significantly faster. Such efficiency is critical for attack detection systems like \toolname as they are resource-constrained~\cite{shaofeiempirical}. Noting that we use large language models for offline \ac{gioc} extraction in Section~\ref{build} due to their superior ability in identifying diverse and nuanced natural language entities. In contrast, we avoid using LLMs to directly analyze system events for efficiency.

\noindent\textbf{Query Acceleration.}
\label{accela}
A naive approach to realizing $ProvQ(e.y, {\theta}_q)$ is to compare $e$ to each \ac{atie}, $t$, in \ac{amid} and then further enumerate each \ac{gioc} within $t$ to calculate the score. However, such a native method is time-consuming and unsuitable for online \ac{apt} attack detection. To accelerate the queries, we devised a two-stage searching method to eliminate the need to compare every \ac{gioc} in \ac{amid}. The key idea is driven by the observation that \acp{gioc} can be very different semantically. Thus, we do not need to compare system events to \acp{gioc} that are semantically far away.  For example, we do not need to calculate the similarity between ``remove'' and ``add a file''. Our two-stage searching method first clusters \acp{gioc} using the Mean-Shift algorithm~\cite{10.1145/1143844.1143864,4270253} based on their embedding vectors. We chose Mean-Shift because it is a non-parametric algorithm that does not require setting hyper-parameters, such as the number of clusters manually. Additionally, it is suitable for distributions with irregular shapes and varying densities, making it capable of handling clusters of arbitrary shapes, which aligns with our scenario where clustering involves a large number of security-specific terms. Thus, during the search, given a field $e.y$ of a system event  $e$, \toolname first identifies the cluster closest to $e.y$ and then finds the most similar \acp{gioc} within that cluster. This approach enables \toolname to avoid comparing $e.y$ with \acp{gioc} in other clusters, which are semantically far away. 

\noindent\textbf{Threshold Setting.}
We employ Grubbs' Test~\cite{couderc2007grubbs}, a standard statistical method for outlier detection, to automatically determine the value of $\theta_{q}$ based on benign data. We opt for this method due to its elimination of the need for manually specified parameters. Specifically, for a given benign dataset comprising system events, we compute the similarity score of each system event within the benign dataset against each \ac{atie} in the \ac{amid}. Subsequently, we apply the one-sided Grubbs' Test to determine the outlier detection threshold for these similarity scores, which serves as $\theta_{q}$.

 \begin{figure*}[!t]
     \centering
     \includegraphics[width=1.02\linewidth]{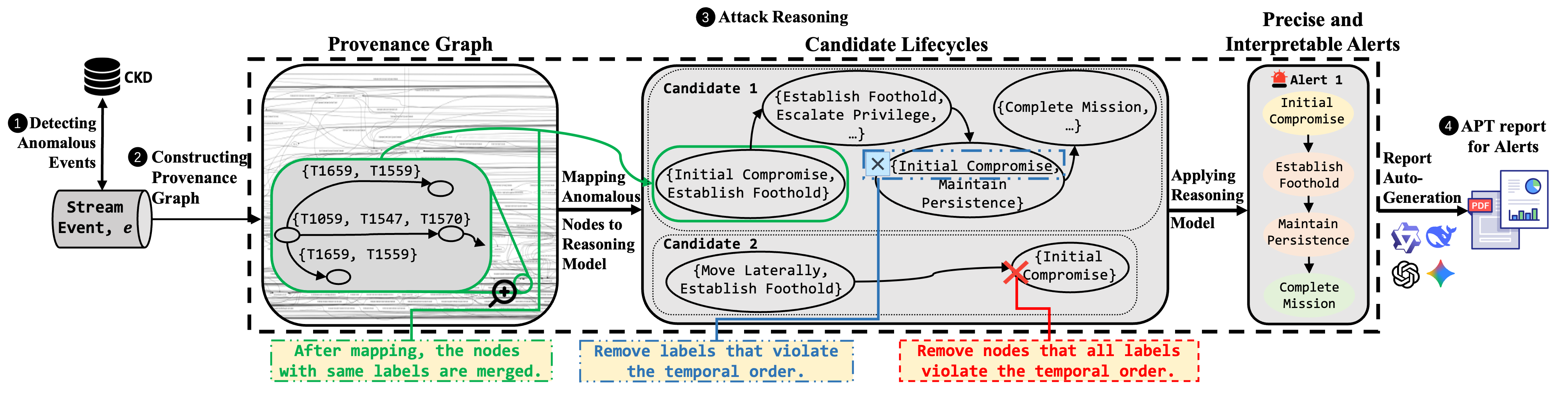}
     \caption{The workflow of \toolname. Given an event stream, \toolname first queries \ac{amid} to detect anomalous events by labeling attack techniques associated with them. Each event may correspond to multiple techniques. Next, \toolname constructs the provenance graph, with a zoomed-in view provided in the figure, highlighting the detected technique labels on the edges.
     Then, \toolname performs attack reasoning on the provenance graph to generate alerts. 
     Specifically, it maps the anomalous nodes to reasoning model based on the technique labels of their outgoing edges, generating candidate lifecycles. After that, \toolname applies the reasoning model to eliminate false positives and produce precise, interpretable alerts.
     Finally, \toolname uses LLM to generate the APT report for each alert, enabling security analysts to rapidly understand the nature of threats and respond effectively.
     }
 	\label{fig:reasoning}
 \end{figure*}

\section{Design of \toolname}
\toolname offers online, accurate, and interpretable detection of APT attacks based on \ac{amid}. The key idea is to utilize a similarity score between system events and \acp{atie} as an indicator of attacks, rather than relying on statistical anomalies detected through data-driven approaches. This design presents two key advantages over data-driven methods. Firstly, the alerts generated by \toolname are more interpretable. Instead of simply presenting statistical anomaly scores, \toolname can explain why a specific event is deemed malicious by identifying the attack technique used at each stage. Secondly, \toolname offers higher precision by reducing false positives based on attack logic, which is a significant limitation in data-driven approaches.

\subsection{Detection Workflow}

The detection workflow of \toolname is illustrated in Figure~\ref{fig:reasoning}. It consists of four steps: \circleblacknum{1}Detecting Anomalous Events: Given a stream of system events, \toolname first matches each incoming event $e$ with \ac{amid} to identify anomalous events. It detects anomalies by querying a system event in \ac{amid} using $ProvQ$, determining whether it corresponds to an attack technique documented in \ac{amid}. If an event matches at least one technique, it is classified as anomalous.
\circleblacknum{2}Constructing Provenance Graph: \toolname constructs a provenance graph based on the detected anomalous events and treats it as a potential seed alert for attacks in an online manner, adopting the same online graph construction approach as \nodlink~\cite{nodlink}, which we chose for its high efficiency. During the graph construction, the event-level anomalies are treated as seed nodes and expand the subgraph by propagating anomaly scores. Based on the attack aggregation assumption~\cite{nodlink}, this expansion introduces lower-scored (benign in event-level detection) nodes, reducing false negatives.
\circleblacknum{3}Applying Reasoning Model: \toolname  analyzes provenance graph with our reasoning model, extended from APT Lifecycle model~\cite{holmes,mandiant}. This step helps eliminate false positives that do not contain sufficient steps to ensure a successful attack or that do not follow the logical sequence of attacks. \circleblacknum{4} Automatically Generating APT Reports: \toolname leverages the lifecycle-styled attack alerts and employs a LLM to generate comprehensive APT reports. This step enables security practitioners to develop an intuitive and holistic understanding of detected attack incidents, including critical information such as \acp{ioc}, attack techniques and behaviors, and recommended mitigation strategies, thereby supporting timely analysis and effective incident response.

Having well described Step \circleblacknum{1} and \circleblacknum{2} above, we now focus on the subsequent steps of the workflow. In Section~\ref{5-b}, we detail the design and usage of the attack reasoning model used in Step \circleblacknum{3}. Section~\ref{5-c} presents how \toolname leverages this model to achieve precise and alert generation. Finally, Section~\ref{5-d} elaborates on the methodology for generating a comprehensive APT report for each alert in Step \circleblacknum{4}.

\subsection{Attack Reasoning}
\label{5-b}
\toolname utilizes the knowledge of the APT Lifecycle to analyze attack graphs and reason about concise, APT Lifecycle-structured attack alerts. The core idea is that a successful attack must include sufficient steps that follow a logical sequence, for example,``Initial Access" must precede data theft. Missing critical stages (e.g., ``Command and Control") suggests an incomplete attack, while steps violating the expected order are less likely to represent genuine threats, helping prioritize high-confidence, logically consistent alerts.

The attack reasoning consists of two steps. First, \toolname maps nodes to the reasoning model based on the industry-standard APT Lifecycle stage models~\cite{tacticweb,mandiantcti}, widely adopted in production environments~\cite{10.1145/3571726, shaofeiempirical}. Second, \toolname uses the reasoning model to verify the temporal order of stages within candidate lifecycles, remove invalid stages, check the completeness of attack stages, and generate alerts. 

\noindent\textbf{Reasoning Model.} 
\toolname's reasoning model is designed with two goals: (1) to provide strong guidance that improves attack detection accuracy, and (2) to ensure high interpretability, enabling security experts to understand detection results without extensive training.

\begin{figure}[!t]
    \centering
    \includegraphics[width=0.98\linewidth]{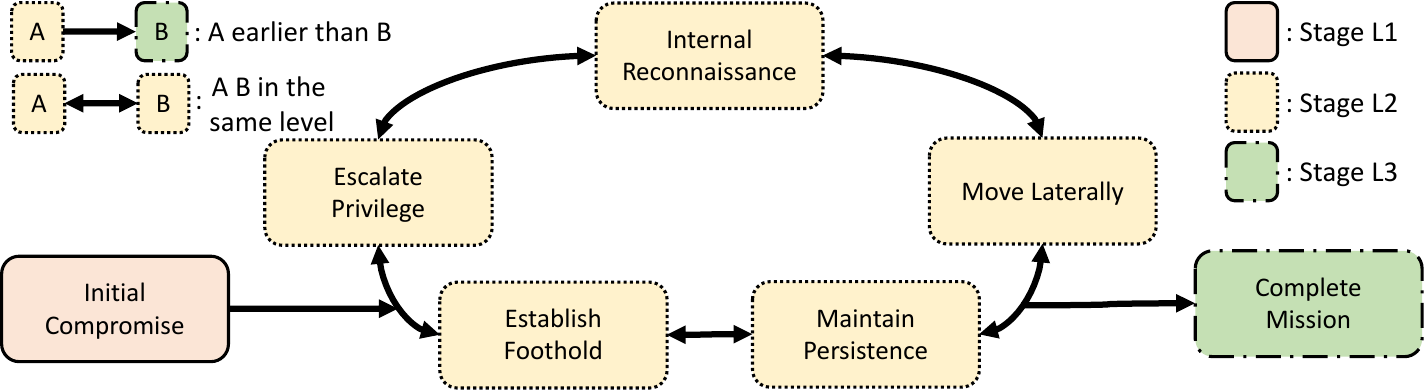}
	\caption{\eat{The temporal order of stages in the APT Lifecycle model. Stages are divided into three levels\eat{, from L1 to L3,} according to their temporal order. The stages connected by double arrows in the middle belong to the same level and possess no inherent order within the attack campaign. The stages linked by single arrows signify sequential attack actions within the attack campaign.}Stages are grouped into three levels based on the temporal order. Double arrows indicate stages within the same level, with no fixed order; single arrows denote sequential progression between stages.}
	\label{fig:killchain}
\end{figure}

To achieve our design goals, we propose a relaxed  APT Lifecycle model based on the widely adopted APT Lifecycle model within the industry ~\cite{lifecyclecrowded,lifecycleibm, lifecycledell,lifecyclemandiant}, shown in Figure~\ref{fig:killchain}, where $A \rightarrow B$ indicates that stage $A$ precedes stage $B$. This model fulfills the first design goal by encapsulating the general understanding of \ac{apt} attacks. It captures the key steps and the temporal relationships among them for typical \ac{apt} attacks in practice, enabling the identification of incomplete or chronologically inconsistent alerts as potential false positives. 
Regarding the second design goal, this model is derived from the standard industrial models, enabling security experts to grasp its terminology without specialized training. 

Note that our reasoning model defines and uses a relaxed APT Lifecycle that only requires the ``\textit{Initial Compromise}'' stage to precede all others and       ``\textit{Complete Mission}'' to occur last. No strict temporal order is imposed on intermediate stages such as ``\textit{Escalate Privilege}''. This design reflects practical observations: attackers must first compromise a system before performing follow-up actions. We relax ordering constraints in the original Lifecycle models for other stages because they often share mutiple techniques—e.g., a C2 mechanism may support both ``\textit{Establish Foothold}'' and ``\textit{Escalate Privilege}''. Relaxing these constraints helps mitigate ambiguity and improves robustness in real-world scenarios.

\noindent\textbf{Usage of the Reasoning Model.}  \toolname uses the reasoning model to reason the candidate lifecycles and infer accurate attack alerts.
It first streamlines alerts by removing stage labels that violate the model’s temporal order. Then, by evaluating lifecycle completeness, it generates alerts for valid sequences and discards incomplete ones—likely false positives from benign behavior.

\textit{Streamlining Alerts.} 
As a highly automated system, \toolname may generate false positives in flagging anomalous events, for instance, misclassifying a benign port scan as ``\textit{Initial Compromise}''. To address this, \toolname removes events that violate the temporal order shown in Figure~\ref{fig:killchain}. For example, as illustrated in Figure~\ref{fig:reasoning}, if \textit{Initial Compromise}'' appears after ``\textit{Establish Foothold}'' in the provenance graph, the former is pruned, as re-compromising the system after already gaining access is logically inconsistent

 \textit{Raising Alerts.}  Our insight to raise alerts is that failed or incomplete attack candidates lacking essential attack behaviors can be mistaken for benign behaviors. For example, an IT maintainer might also exhibit behavior resembling port scanning during the ``\textit{Initial Compromise}'' stage. Simply detecting these incomplete attacks can lead to false positives. Therefore, \toolname necessitates that an alert encompasses at least the ``\textit{Initial Compromise}'' and ``\textit{Establish Foothold}'' stages, as they are essential for an attacker to penetrate the victim system. Additionally, \toolname demands that the alert include at least one stage from ``\textit{Escalate Privilege}'', ``\textit{Internal Reconnaissance}'', ``\textit{Move Laterally}'', and ``\textit{Maintain Persistence}'' because these represent the essential behaviors in an attack. Notice that the presence of ``\textit{Complete Mission}'' is optional, as not all attackers will erase their tracks or harm the victim's system. 

 \subsection{Mapping Anomalous Nodes to the Reasoning Model}
 \label{5-c}
 To enable reasoning, \toolname needs to map anomalous nodes in alerts to stages within our reasoning model to generate the candidate lifecycles. To this end, we first map the technique labels of each anomalous node to a tactic label. This step is straightforward, as the ATT\&CK framework already provides the mapping relations~\cite{attckweb}. Then, we employ the mapping rules in Table~\ref{tab:tac2stage-appendix}, derived from our comprehension of the APT Lifecycle model, to assign the tactic labels to the respective APT Lifecycle stages. %These rules are summarized based on the ATT\&CK framework documentation.

\begin{table}[h!]
\centering

\caption{Tactic-Lifecycle Stage Mapping Table.}
\label{tab:tac2stage-appendix}
\resizebox{0.95\linewidth}{!}{%
\begin{tabular}{cc}
\toprule
\textbf{Tactic}               & \textbf{APT Lifecycle Stage}       \\ 
\toprule
Reconnaissance, Initial Access       & Initial Compromise     \\ \hline
\multirow{2}{*}{\begin{tabular}[c]{@{}c@{}}Execution, Resource Development,\\  Command and Control \end{tabular}} & \multirow{2}{*}{Establish Foothold} \\    &    \\ \hline
Privilege Escalation, Credential Access & Escalate Privilege     \\ \hline
Discovery, Collection             & Internal Reconnaissance \\ \hline
Lateral Movement     & Move Laterally         \\ \hline
Persistence, Defense Evasion          & Maintain Persistence      \\ \hline
Exfiltration, Impact         & Complete Mission      \\ \bottomrule
\end{tabular}%
}
\end{table}

The key challenge in mapping anomalous nodes to stage labels lies in the one-to-many relationships: a node may have multiple edges, while each single edge may correspond to multiple \acp{atie} in \ac{amid}, linked to different tactics and thus multiple APT stages. For instance, an alarm edge involving ``T1053 Scheduled Task" and ``T1546 Event Triggered Execution" may relate to tactics like ``Execution", ``Persistence" and ``Privilege Escalation", mapping to stages including ``Maintain Persistence" and ``Escalate Privilege". As nodes typically have many outgoing edges, aggregating edge-level stage labels directly would assign multiple stages per node, causing a path explosion during reasoning due to the need to evaluate all stage label combinations for temporal order.

To address the one-to-many mapping problem, we transform it into a one-to-limited mapping via label merging, assigning edges, and subsequently nodes, a limited set of stage labels. Our core concept is to retains stage labels with ``relatively high'' scores, computed as the cumulative similarity ($Sim(e, t)$) between event $e$ and matching \acp{atie}. Using Grubbs' Test, we identify high outliers as ``relatively high" scores and keep their corresponding stages; if no high outliers exist, we remove low outliers and retain the rest. This strategy avoids over-reliance on the single highest-scoring stage, which may be a false positive. Since node stage labels are derived from their outgoing edges, which also yields a one-to-many mapping, we apply the same one-to-limited method to assign the node a refined set of stage labels with ``relatively high'' scores.

Specifically, We deduce the correct stage labels for a node by performing one-to-limited mapping over its outgoing edges. Nodes without successors, which means they have no generated events or outgoing edges, are skipped, and stage selection begins from their predecessors. For node $n$, the stage labels of its outgoing edges serve as candidate options, each weighted by its corresponding anomaly score. The labels with ``relatively high" weights (determined via Grubbs' Test, as before) are selected for $n$. We then backtrack to $n$'s predecessor to vote on its stage labels, repeating this process iteratively until all nodes in the attack graph are assigned APT lifecycle stages. Finally, nodes sharing the same stage label are merged into single nodes in the reasoning model to form candidate lifecycles, as illustrated in Figure~\ref{fig:reasoning}.

\subsection{APT Report Generation}
\label{5-d}
\toolname leverages a large language model (LLM) to generate comprehensive and human-readable APT reports for detected alerts. This capability enables security analysts to rapidly understand the nature of threats and respond effectively.

\eat{
The core insight behind this design is that the APT report should be the primary form of the alert, not a post-hoc artifact. By treating alerts as communication tools for humans, we elevate them from technical signals to decision-supporting artifacts.}

The core insight behind this design is that the form of an alert fundamentally affects its utility in security operations.
Traditional alerting systems either produce isolated \acp{ioc}\cite{rhoades2014machine, ttpdrill}, which lack context, or generate low-level provenance graphs in which nodes represent system entities such as processes and files and edges represent system calls\cite{nodlink, kairos, magic, trec, holmes}. While these representations capture technical details, they are insufficient for effective human analysis. Analysts must operate with critical information gaps: there is typically no high-level overview to help prioritize alerts, no chronological event flow to clarify attack progression, and no mitigation guidance to support timely response. Therefore, a more effective alert format should directly provide this missing context.

Based on this insight, we propose a new alerting paradigm: the structured, analyst-centric APT report should serve as the primary form of the alert, replacing low-level system-centric representations such as provenance graphs. 
To realize this paradigm, we draw upon real-world APT reports published by leading cybersecurity organizations~\cite{apt37fireeye,apt42fireeye,apt43mandiant}, whose effectiveness in communicating complex threats stems from their consistent organization into three complementary content dimensions: high-level context, technical details, and actionable guidance. 
High-level context provides a concise overview of the attack, including the likely threat actor, their objectives, and the potential business impact. This type of information is essential for rapid triage and strategic decision-making by security managers. Technical details supports in-depth investigation by presenting evidence such as attack timelines, system-level provenance graphs, attack lifecycle and \acp{ioc}. These elements allow analysts to reconstruct the sequence of malicious activities and validate detection results. actionable guidance offers concrete actions that defenders can take, such as terminating malicious processes, blocking suspicious network connections, or deploying new detection rules. Such recommendations reduce response latency and improve the consistency of incident handling. 
This tripartite structure ensures that the report serves diverse stakeholders across different phases of incident response.

To generate these components, we adopt a conditional synthesis strategy that integrates the output of attack reasoning with LLM synthesis. For the technical details, we directly utilize the attack provenance graph, the attack lifecycle and the extracted IoCs produced by earlier stages. This ensures that the report is grounded in accurate, system-observed data and maintains traceability to the underlying detection evidence. For the high-level context and actionable guidance components, which require higher-level synthesis and contextual interpretation, we employ an LLM to generate analyst-centric narratives from the structured lifecycle-styled alerts. 

However, a key challenge lies in ensuring that the LLM-generated content is not only fluent but also factually faithful, structurally consistent, and operationally actionable. Without proper guidance, LLMs may introduce hallucinated entities or events, omit critical information, or produce recommendations that lack executable specificity, rendering the output unreliable for security operations. 

To address this, we propose a methodology for evidence-constrained content generation. This approach systematically aligns LLM-generated content with observed attack data through three complementary mechanisms:

\textbf{Controlled Semantic Mapping.}  
We first define formal schemas for high-level context and actionable guidance, specifying the required information types (e.g., threat actor, business impact, mitigation steps). These schemas are also paired with curated exemplars from real-world APT reports in our knowledge sources in Section~\ref{eval-a}, enabling the LLM to learn the expected semantic structure and abstraction level. This mapping ensures the consistency of the LLM in content scope and reporting style.

\textbf{Structured Generation Prompt.}
Next, we orchestrate the generation process using a structured prompt that enforces a three-part control mechanism:  
(1) a clear role definition (e.g., ``you are a security analyst''),  (2) a step-by-step generation procedure that decomposes the task into intent inference, impact assessment, and recommendation formulation, and  (3) domain-specific constraints that require the model to base its output strictly on the provenance graph, avoid speculation, and format results in a consistent, machine-readable structure. A prompt sample used in \toolname is presented in Figure~\ref{prompt:llm-guidance}.

\begin{figure}[t]
\begin{promptbox}{LLM Prompt for Generating High-Level Context and Actionable Guidance}
You are a cybersecurity analyst generating an alert report based on observed system activity. Use only the evidence provided below.

## Evidence:
**Alert Lifecycle**: ...
**Attack Behaviors in Each Stage of Lifecycle with the Timestamp**: ...

## Generating contents:
**High-Level Context**: Summarize the attack in one paragraph. Include:
- Threat actor (if identifiable)
- Likely objectives
- Business impact

**Actionable Guidance**: Provide specific recommendations. For each, include:
- Mitigation step
- Detection rule (e.g., YARA, Sigma)
- Hardening strategy

Do not invent information. All claims must be supported by the provided evidence. 
\end{promptbox}
\caption{Structured prompt used to guide the LLM in generating high-level context and actionable guidance.}
\label{prompt:llm-guidance}
\end{figure}

\textbf{Post-Hoc Factual Verification.} 
To further enhance reliability, we incorporate factual validation mechanisms. First, during generation, the LLM is constrained to reference only entities and events that appear in the attack provenance graph, ensuring factual faithfulness. Second, after generation, we apply a verification step in which the LLM cross-checks its own output against the original detection data using a validation prompt. A typical example is shown in Figure~\ref{prompt:llm-validation}.

\begin{figure}[]
\begin{promptbox}{LLM Prompt for Validating the Generated Content}
You are a cybersecurity analyst validating the authenticity of the content of an alert report based on observed system activity. 

## Evidence:
**Alert Lifecycle**: ...
**Attack Behaviors in Each Stage of Lifecycle with the Timestamp**: ...

Review the following actionable guidance from the alert report: "Block C2 server at 192.168.1.100."  
Verify whether this IP address relates to events of the provided evidence. 

If not, mark it as unsupported.
\end{promptbox}
\caption{Validation prompt used to verify the factual support.}
\label{prompt:llm-validation}
\end{figure}

Together, these three mechanisms form a comprehensive content assurance strategy: 
controlled semantic mapping establishes the expected structure, structured prompting enforces evidence adherence during generation, and post-hoc verification detects residual inaccuracies. 
This end-to-end control ensures that the final APT report is analyst-ready and operationally reliable.

In this way, \toolname transforms low-level detection results into coherent, analyst-centric APT reports.

\section{Evaluation}\label{section5}
We implemented a prototype of \toolname and evaluated it with realistic attack scenarios. In this section, we first detail the implementation and the experimental environment. Then, we introduce our evaluation protocol, including metrics, knowledge sources, datasets, and baselines. To evaluate \toolname, we focus on answering the following research questions: 

\noindent \textbf{RQ 1}: Can \toolname detect \ac{apt} attacks accurately and precisely with the help of \ac{gioc}?

\noindent \textbf{RQ 2}: Does \toolname correctly label the ATT\&CK techniques to system events?

\noindent \textbf{RQ 3}: Is \toolname efficient enough for online detection?

\noindent \textbf{RQ 4}: Is \toolname robust against mimicry attacks, incomplete attack lifecycles, new and unseen attacks?

\noindent \textbf{RQ 5}: How each component in \toolname contributes to the overall detection process?

\subsection{Implementation and Experiment Environment}
\label{eval-a}
We implemented \toolname on Python3.8.8 with around 2,000 lines of code (LoC) across all components. In our deployment, \toolname utilizes the FastText model~\cite{fasttext} to encode \acp{gioc} and key information from input system events. The embedding space of FastText is pre-trained through CBOW model~\cite{2013Efficient} using over 10,000 attack-related statements extracted from 1,500 paragraphs in the MITRE ATT\&CK technique description~\cite{attckweb} and 80 pieces of attack descriptions sourced from public CTI reports~\cite{threatminer}. Meanwhile, \toolname employs Sklearn~\cite{sklearn} to implement the Mean-Shift algorithm in the query acceleration component (Section~\ref{accela}). For \ac{amid} and dataset management, \toolname leverages the ElasticSearch database~\cite{elasticsearch2018elasticsearch} for data management and querying. All experiments were conducted on a Ubuntu 22.04 machine with a GTX 3090 GPU, a 40-core 2.40GHz CPU, and 128GB of main memory.

\subsection{Evaluation Protocol}
\label{protocol}
\noindent\textbf{Metrics.} NodLink~\cite{nodlink} has proposed to use graph-level and node-level accuracy to evaluate the effectiveness of attack detection more comprehensively. 
Graph-level accuracy comprises two parts: graph-level precision and graph-level recall, which are defined as $\frac{GTP}{GTP+GFP}$ and $\frac{GTP}{GTP+GFN}$, respectively. $GTP$, $GFP$, and $GFN$ stand for graph-level true positives, false positives, and false negatives, respectively. 
The graph-level positive is the number of reported graphs that contain attack steps, while the negative represents the number of reported graphs that do not contain attack steps.
Node-level accuracy also comprises two parts: node-level precision and node-level recall,
which are defined as $\frac{NTP}{NTP+NFP}$ and $\frac{NTP}{NTP+NFN}$, respectively. $NTP$, $NFP$, and $NFN$ are the numbers of node-level true positives, false positives, and false negatives, respectively. 
Given a reported provenance graph, we consider a node in the graph as an $NTP$ if it represents an attack step. 

\noindent\textbf{Knowledge Sources.} To build \ac{amid}, \toolname crawled the technique descriptions from ATT\&CK v10.0, which encompassed 567 techniques. Additionally, 80 \ac{cti} reports containing attack-related sentences were randomly selected from ThreatMiner~\cite{threatminer} and Mandiant~\cite{mandiantcti}. Following the construction of \ac{amid}, \toolname generated one \ac{atie} for each technique, amounting to a total of 567 entries. Among these, there were 27,652 \acp{gioc} and 1,795 \acp{ioc}. All our \ac{cti} reports are dated before December 2022.

\noindent\textbf{Datasets.} The general information of our datasets is shown in Table~\ref{tab:datasets}. We first evaluate \toolname on three public datasets: THEIA, TRACE, and In-lab Arena. 
THEIA and TRACE originate from the DARPA Transparent Computing Engagement \#3 (E3) database~\cite{darpatc}, which has been extensively utilized in recent researches~\cite{nodlink,kairos, shadewatcher, flash}. And we labeled the ground truth nodes following the E3 document~\cite{e3-doc}. We did not incorporate the CADET dataset  because the necessary information, such as full command lines and parameters, is missing in it, hindering fine-grained knowledge discovery. 
In-lab Arena~\cite{simu} was simulated by \nodlink~\cite{nodlink}, replicating real attacks that took place at Sangfor, one of the most prominent Chinese security vendors. We rely on the documents provided by the \nodlink repository~\cite{nodlink} for labeling the attack techniques in In-lab Arena. 
We refrain from using the DARPA Transparent Computing Engagement \#5 (E5) dataset because we observed discrepancies between attack reports and actual logs in several cases in it. For example, it lacks the corresponding write event of the ``sshdlog'' injection operation documented in THEIA.

\begin{table}[t!]
  \caption{Summary of our evaluation datasets. ``Duration'' and ``Event Rate'' denote the duration of data collection and the average number of events generated per second, respectively.}
  \label{tab:datasets}
      \resizebox{0.49\textwidth}{!}{
  \begin{tabular}{|c|c|c|c|c|c|}
  \hline
  \textbf{Dataset} & \textbf{\# \acp{apt}} & \textbf{Duration} & \textbf{\# Hosts} & \textbf{Event Rate} & \textbf{\# Attack Actions} \\ \hline
  %CADETS           & 3                   & 247h              & 1               & 16.87 eps           & 21                             \\ \hline
  THEIA            & 1                   & 247h              & 1               & 11.25 eps           & 97                             \\ \hline
  TRACE            & 2                   & 264h              & 1               & 75.76 eps           & 93                             \\ \hline
  In-lab Arena    & 5                   & 144h              & 5               & 48.23 eps           & 202                            \\ \hline
  NewlySim        & 2                    & 336h                  &  3               & 168.69 eps    & 39                               \\ \hline
  Open-World & 6 & 168h & 186 & 28.13eps & 212 \\ \hline
  \end{tabular}}
  \end{table}

In addition to these three datasets, we constructed a newly simulated dataset, NewlySim, for our evaluation, with the aim of testing the effectiveness of \toolname in handling unknown new attacks. Specifically, using NewlySim, we exploited two new vulnerabilities, CVE-2023-22809~\cite{cve-2023-22809} and CVE-2024-28085~\cite{cve-2024-28085}. 
These vulnerabilities were selected for their high risk and broad impact. For example, CVE-2023-22809 is rated ``High” by the NVD~\cite{nvd}, affects multiple versions of ``sudo", posing a widespread threat. Similarly, CVE-2024-28085 targets util-linux, which is a core component in most Linux distributions, making it highly relevant to Linux users. Most importantly, neither exploit is covered in our \acp{cti}, as our \ac{amid} contains only \acp{cti} published before December 2022. 

To build NewlySim, we deployed a three-machine scenario comprising a hijacked attacker machine, an extranet interactive jump server machine, and an intranet working machine. The adversary first gained access to the intranet interactive jump server machine through social engineering channels and then utilized the "ssh" service on the springboard to infiltrate the intranet machines. On the intranet working machine, the adversary launched two attack campaigns, leveraging the aforementioned vulnerabilities to achieve privilege escalation. With system-level privileges, the adversary executed various malicious actions that are common in APT attacks, including malicious payload acquisition and execution \cite{anssi, apple}, internal environment reconnaissance \cite{slowik, bazar}, OS credential dumping \cite{minik, rusu}, important file scanning and collection \cite{rocket, black}, information leakage \cite{mstic, great}, and attack evidence eradication \cite{smith, mar}, thus completing an APT attack. We also used a trusted machine to log into the intranet machine to generate benign data through regular Microsoft Office operations. The attack lasted for 168 hours, during which we also collected the benign dataset in the same environment for 168 hours.

Finally, we also evaluated \toolname using an Open-World dataset generated from a realistic industrial environment. We deployed \toolname within one of the world's largest cloud providers, Huawei, and participated in their internal penetration test. During our experiment, \toolname was deployed across over 180 endpoints, including diverse servers, workstations, and desktops, all running various operating systems such as Linux, Windows and openEuler. We first collected benign data over a 20-hour period and manually verified it to ensure no attacks were present. Then, a professional red team initiated attacks on the machines monitored by \toolname over a seven-day period. The ground truth for the Open-World dataset was provided by the this red team.

For the DARPA datasets, we follow the same practice of NodLink~\cite{nodlink} to split the training and test datasets. We first select 20\% of the data that covers the time period of attacks as the test dataset. Then, we use the remaining 80\% of the data, which only contains benign events, as the training dataset~\cite{nodlink}. For other datasets, we use the provided benign data for training and anomaly data for testing. We first apply event-level detection on the training set to derive the threshold using Grubbs' Test, and then evaluate \toolname’s performance on the test set.

\noindent\textbf{Overlap Between \ac{cti} Reports and Data.} 
To ensure the fairness of the evaluation, we have verified that the attack techniques utilized in our datasets are NOT documented in our \ac{cti} reports directly. For the DARPA dataset, we manually inspected it to confirm that none of the knowledge in our \ac{cti} reports were used in the attacks. For the In-lab Arena dataset and the Open-World dataset, all of our \ac{cti} reports are dated before December 2022, which is before the generation dates of these datasets. We further examined the attack techniques employed in these two datasets and confirmed that they were not mentioned in the \ac{cti} reports. In our NewlySim dataset, we deliberately chose to use techniques that are dated after January 2023.

\noindent\textbf{Baselines.} To evaluate the effectiveness of \toolname in detecting APT attacks, we compare it with five end-to-end provenance-based APT attack detection systems: \holmes~\cite{holmes},  \airtag~\cite{airtag}, \nodlink~\cite{nodlink}, \kairos~\cite{kairos}, and \extractor~\cite{extractor}. 
The first four baselines are non-\ac{cti} driven detection systems and \extractor is a state-of-the-art \ac{cti}-driven system that extracts \ac{ioc} patterns from \ac{cti} reports. 
\extractor is a representative CTI-extraction system that extracts \ac{ioc} graphs from \ac{cti} reports. Following the evaluation methodology in its paper~\cite{extractor}, we use \extractor to generate graphs and feed them into \poirot for detection, ensuring a fair and comparable evaluation.
To evaluate the effectiveness of \ac{gioc} in \ac{apt} attack detection, we replace the knowledge in \ac{amid} with knowledge extracted using two state-of-the-art knowledge extraction methods: \ttpdrill~\cite{ttpdrill} and \ladder~\cite{ladder}. We also compare \toolname with an \ac{ioc}-only version of \toolname, which utilizes \acp{ioc} extracted by IoCParser in the knowledge base. Table~\ref{tab:baseline} lists a summary of all the baseline systems. These three CTI extraction baselines are designated as ``\ttpdrill + \toolname'', ``\ladder + \toolname'', and ``\toolname with only \ac{ioc}''.

\noindent\textbf{Setup for Baselines.} For \airtag, \nodlink, \kairos, \extractor, \ttpdrill, and \ladder, we utilize the open-source code released by the respective authors. We apply all the optimizations and recommended hyperparameters outlined in their original papers to train the models using the same benign dataset as \toolname. For \holmes, we adopt the implementation provided by the authors of \nodlink. For \airtag, we retrain the BERT using the benign dataset, which is identical to the one used for the other baselines and \toolname, following the training parameters specified in their open-source code.  

\begin{table}[]
 \caption{Summary of our baselines.} 
\label{tab:baseline}
\centering
\resizebox{\linewidth}{!}{
\begin{tabular}{|c|c|c|c|c|}
\hline
\textbf{Type} & \textbf{Name}      & \textbf{Methodology}     & \textbf{Venue} & \textbf{Year} \\ \hline
\multirow{6}{*}{\begin{tabular}[c]{@{}c@{}} Detection\\Baselines\end{tabular}} & HOLMES    & Manual Rules     &  IEEE S\&P   & 2019 \\ \cline{2-5}
 & AIRTAG & Date-driven & Usenix Security & 2023   \\ \cline{2-5}
 & NodLink   & Date-driven &    NDSS   &   2024   \\ \cline{2-5} 
 & KAIROS    & Date-driven &  IEEE S\&P    &    2024  \\ \cline{2-5}
 & \multirow{2}{*}{\begin{tabular}[c]{@{}c@{}}EXTRACTOR\\+ POIROT\end{tabular}} & \multirow{2}{*}{CTI-IoC-driven} & \multirow{2}{*}{Euro S\&P} & \multirow{2}{*}{2021} \\ 
 & & & & \\ \hline
\multirow{6}{*}{\begin{tabular}[c]{@{}c@{}}\ac{cti}\\Extraction\\Baselines\end{tabular}} 
 & \multirow{2}{*}{\begin{tabular}[c]{@{}c@{}}TTPDrill +\\\toolname*\end{tabular}} & \multirow{2}{*}{\begin{tabular}[c]{@{}c@{}}Basic Syntax\\Analysis\end{tabular}} & \multirow{2}{*}{ACSAC} & \multirow{2}{*}{2017} \\
 & & & & \\ \cline{2-5}
 & \multirow{2}{*}{\begin{tabular}[c]{@{}c@{}}LADDER +\\\toolname*\end{tabular}} & \multirow{2}{*}{Language Model} & \multirow{2}{*}{RAID} & \multirow{2}{*}{2023} \\
 & & & & \\ \cline{2-5}
 & \multirow{2}{*}{\begin{tabular}[c]{@{}c@{}}\toolname \\with only \ac{ioc}\end{tabular}} & \multirow{2}{*}{\begin{tabular}[c]{@{}c@{}}Regular Expression \\matching\end{tabular}} & \multirow{2}{*}{NA} & \multirow{2}{*}{NA} \\
 & & & & \\ \cline{2-5}
 \hline
                                                                                       
\end{tabular}}
\end{table}

\begin{table}[]
\centering
\caption{The graph-level accuracy results for \toolname and baselines. P stands for precision, and R stands for recall. }
\label{tab:graph-level}

\begin{adjustbox}{max width=\linewidth}
\large % 或 \footnotesize 等，控制字体大小
\begin{tabular}{|c|cc|cc|cc|cc|cc|}
\hline
 & \multicolumn{10}{c|}{\textbf{Dataset}} \\ \hline
\multirow{2}{*}{\begin{tabular}[c]{@{}c@{}}Detection\\Baselines\end{tabular}} &
  \multicolumn{2}{c|}{THEIA} &
  \multicolumn{2}{c|}{TRACE} &
  \multicolumn{2}{c|}{In-Lab Arena} &
  \multicolumn{2}{c|}{NewlySim} &
  \multicolumn{2}{c|}{Open-World}\\ \cline{2-11}
 &
  \multicolumn{1}{c}{P} &
  R &
  \multicolumn{1}{c}{P} &
  R &
  \multicolumn{1}{c}{P} &
  R &
  \multicolumn{1}{c}{P} &
  R &
  \multicolumn{1}{c}{P} &
  R \\ \hline
\textbf{\holmes} &
  \multicolumn{1}{c}{\textbf{1.00}} &
  \textbf{1.00} &
  \multicolumn{1}{c}{0.15} &
  \textbf{1.00} &
  \multicolumn{1}{c}{0.04} &
  \textbf{1.00} &
  \multicolumn{1}{c}{0.14} &
  \textbf{1.00} &
  \multicolumn{1}{c}{0.15} &
  0.40 \\ \hline
\textbf{\airtag} &
  \multicolumn{1}{c}{\textbf{1.00}} &
  \textbf{1.00} &
  \multicolumn{1}{c}{\textbf{1.00}} &
  \textbf{1.00} &
  \multicolumn{1}{c}{0.83} &
  \textbf{1.00} &
  \multicolumn{1}{c}{1.00} &
  \textbf{1.00} &
  \multicolumn{1}{c}{0.63} &
  \textbf{1.00} \\ \hline
\textbf{\nodlink} &
  \multicolumn{1}{c}{\textbf{1.00}} &
  \textbf{1.00} &
  \multicolumn{1}{c}{0.67} &
  \textbf{1.00} &
  \multicolumn{1}{c}{\textbf{1.00}} &
  \textbf{1.00} &
  \multicolumn{1}{c}{\textbf{1.00}} &
  \textbf{1.00} &
  \multicolumn{1}{c}{0.71} &
  \textbf{1.00} \\ \hline
\textbf{\kairos} &
  \multicolumn{1}{c}{0.91} &
  \textbf{1.00} &
  \multicolumn{1}{c}{0.88} &
  0.88 &
  \multicolumn{1}{c}{\textbf{1.00}} &
  \textbf{1.00} &
  \multicolumn{1}{c}{0.50} &
  \textbf{1.00} &
  \multicolumn{1}{c}{0.63} &
  \textbf{1.00} \\ \hline
\begin{tabular}[c]{@{}c@{}}\textbf{\extractor}\\ \textbf{+ \poirot}\end{tabular} &
  \multicolumn{1}{c}{0.33} &
  \textbf{1.00} &
  \multicolumn{1}{c}{0.33} &
  \textbf{1.00} &
  \multicolumn{1}{c}{0.50} &
  \textbf{1.00} &
  \multicolumn{1}{c}{0.25} &
  \textbf{1.00} &
  \multicolumn{1}{c}{0.43} &
  0.30 \\ \hline
\textbf{\toolname} &
  \multicolumn{1}{c}{\textbf{1.00}} &
  \textbf{1.00} &
  \multicolumn{1}{c}{\textbf{1.00}} &
  \textbf{1.00}&
  \multicolumn{1}{c}{\textbf{1.00}} &
  \textbf{1.00} &
  \multicolumn{1}{c}{\textbf{1.00}} &
  \textbf{1.00} &
  \multicolumn{1}{c}{\textbf{0.91}} &
  \textbf{1.00} \\ \hline    
\end{tabular}
\end{adjustbox}
\end{table}

\begin{table}[]
\centering
\caption{The node-level accuracy results for \toolname and baselines. P stands for precision, and R stands for recall. }
\label{tab:node-level}

\begin{adjustbox}{max width=\linewidth}
\large % 或 \footnotesize 等，控制字体大小
\begin{tabular}{|c|cc|cc|cc|cc|cc|}
\hline
 & \multicolumn{10}{c|}{\textbf{Dataset}} \\ \hline
\multirow{2}{*}{\begin{tabular}[c]{@{}c@{}}Detection\\Baselines\end{tabular}} &
  \multicolumn{2}{c|}{THEIA} &
  \multicolumn{2}{c|}{TRACE} &
  \multicolumn{2}{c|}{In-Lab Arena} &
  \multicolumn{2}{c|}{NewlySim} &
  \multicolumn{2}{c|}{Open-World}\\ \cline{2-11}
 &
  \multicolumn{1}{c}{P} &
  R &
  \multicolumn{1}{c}{P} &
  R &
  \multicolumn{1}{c}{P} &
  R &
  \multicolumn{1}{c}{P} &
  R &
  \multicolumn{1}{c}{P} &
  R \\ \hline
\textbf{\holmes} &
  \multicolumn{1}{c}{0.01} &
  0.98 &
  \multicolumn{1}{c}{0.01} &
  0.74 &
  \multicolumn{1}{c}{0.01} &
  0.32 &
  \multicolumn{1}{c}{0.01} &
  0.40 &
  \multicolumn{1}{c}{0.04} &
  0.21 \\ \hline
\textbf{\airtag} &
  \multicolumn{1}{c}{0.31} &
  0.84 &
  \multicolumn{1}{c}{0.26} &
  0.88 &
  \multicolumn{1}{c}{0.18} &
  0.96 &
  \multicolumn{1}{c}{0.19} &
  0.86 &
  \multicolumn{1}{c}{0.30} &
  0.87 \\ \hline
\textbf{\nodlink} &
  \multicolumn{1}{c}{0.23} &
  \textbf{1.00} &
  \multicolumn{1}{c}{0.25} &
  \textbf{0.98} &
  \multicolumn{1}{c}{0.17} &
  0.92 &
  \multicolumn{1}{c}{0.28} &
  0.94 &
  \multicolumn{1}{c}{0.48} &
  0.90 \\ \hline
\textbf{\kairos} &
  \multicolumn{1}{c}{0.13} &
  0.93 &
  \multicolumn{1}{c}{0.11} &
  0.94 &
  \multicolumn{1}{c}{0.32} &
  0.92 &
  \multicolumn{1}{c}{0.17} &
  0.96 &
  \multicolumn{1}{c}{0.33} &
  0.94 \\ \hline
\begin{tabular}[c]{@{}c@{}}\textbf{\extractor}\\ \textbf{+ \poirot}\end{tabular} &
  \multicolumn{1}{c}{0.34} &
  0.77 &
  \multicolumn{1}{c}{0.34} &
  0.88 &
  \multicolumn{1}{c}{0.56} &
  0.54 &
  \multicolumn{1}{c}{0.29} &
  0.25 &
  \multicolumn{1}{c}{0.38} &
  0.22 \\ \hline
\textbf{\toolname} &
  \multicolumn{1}{c}{\textbf{0.62}} &
  \textbf{1.00} &
  \multicolumn{1}{c}{\textbf{0.82}} &
  \textbf{0.98}&
  \multicolumn{1}{c}{\textbf{0.82}} &
  \textbf{1.00} &
  \multicolumn{1}{c}{\textbf{0.78}} &
  \textbf{1.00} &
  \multicolumn{1}{c}{\textbf{0.82}} &
  \textbf{1.00} \\ \hline    
\end{tabular}
\end{adjustbox}
\end{table}

\subsection{RQ 1: Effectiveness in APT Attack Detection}
\label{sec:accuracy}
We calculate the graph- and node-level accuracy of \toolname and the baselines to evaluate the effectiveness of \toolname in APT detection. The results are shown in Table~\ref{tab:graph-level} and Table~\ref{tab:node-level}. \toolname performs both the best graph- and node-level accuracies compared to the baselines.  

\noindent\textbf{Graph-Level Accuracy.} For graph-level accuracy, \toolname achieves the precision and recall of 1.00, except for Open-World dataset. There is a false positive of \toolname at the graph level in the Open-World dataset. After an investigation by the red team, it was confirmed that the false positive was caused by an update to the system's security software, which involved serious high-risk behaviors similar to attacks. Nevertheless, all baselines also fail to detect this attack. Data-driven methods (i.e., \nodlink, \kairos, and \airtag) have low graph-level precision because they rely on extracting normal behavior characteristics from benign data, which cannot handle rare but benign events. \holmes's graph-level precision is the lowest due to the incompleteness of the knowledge included and the over-generalization in the handling of objects, such as treating all non-trusted IP addresses as a means of initial compromise.

\noindent\textbf{Node-Level Accuracy.} For node-level accuracy, \toolname has higher node-level precision and recall in each dataset than the baselines.  Most baselines suffer from low precision and recall, because they cannot identify rare but benign nodes or have complete rules capturing all attack behaviors. For data-driven approaches, the training set significantly impacts detection effectiveness. That said, if normal behaviors are not included in the training set, they are prone to being identified as malicious. For example, in NewlySim, a normal user's unique file handling process, including using the \textit{tar} command (absent from benign data), led to false alarms (NFPs) by all data-driven baselines except \nodlink. While \nodlink employs Grubbs' test and limited-step anomaly score propagation to reduce NFPs, it still generates a significant number due to the absence of node behavior and temporal logical order checks, as seen in \toolname. \holmes performs the worst due to its coarse-grained rule design. For example, \holmes categorizes read operations originating from untrusted IP addresses as $Untrusted\_Read$. However, in normal scenarios, numerous legitimate network access operations occur, resulting in a large number of IP addresses that have never appeared in the system becoming untrusted, causing false positives.

\noindent\textbf{Effectivenss of \ac{gioc}.} To further evaluate the effectiveness of \ac{gioc} in \ac{amid}, we compare \toolname's performance augmented with \ac{gioc} knowledge against three CTI extraction baselines. We find that \toolname with \ac{gioc} achieves superior accuracy in both graph- and node-level detection, outperforming all baselines. The results are shown in Table~\ref{tab:graph-level-cti} and Table~\ref{tab:node-level-cti}. 

\begin{table}[h]
\centering
\caption{The graph-level accuracy results for \toolname and baselines. P stands for precision, and R stands for recall. }
\label{tab:graph-level-cti}

\begin{adjustbox}{max width=0.96\linewidth}
\large % 或 \footnotesize 等，控制字体大小
\begin{tabular}{|c|cc|cc|cc|cc|}
\hline
 & \multicolumn{6}{c|}{\textbf{CTI Extraction Baselines}} & \multicolumn{2}{c|}{\textbf{Ours}} \\ \hline
\multirow{2}{*}{\textbf{Dataset}} &
  \multicolumn{2}{c|}{\begin{tabular}[c]{@{}c@{}}\textbf{\ttpdrill +}\\ \textbf{\toolname*}\end{tabular}} &
  \multicolumn{2}{c|}{\begin{tabular}[c]{@{}c@{}}\textbf{\ladder +}\\ \textbf{\toolname*}\end{tabular}} &
  \multicolumn{2}{c|}{\begin{tabular}[c]{@{}c@{}}\textbf{\toolname}\\ \textbf{with only IOC}\end{tabular}} &
  \multicolumn{2}{c|}{\textbf{\toolname}} \\ \cline{2-9} 
 &
  \multicolumn{1}{c}{P} &
  R &
  \multicolumn{1}{c}{P} &
  R &
  \multicolumn{1}{c}{P} &
  R &
  \multicolumn{1}{c}{P} &
  R \\ \hline
THEIA &
  \multicolumn{1}{c}{0.17} &
  \textbf{1.00} &
  \multicolumn{1}{c}{0.20} &
  \textbf{1.00} &
  \multicolumn{1}{c}{0.50} &
  \textbf{1.00} &
  \multicolumn{1}{c}{\textbf{1.00}} &
  \textbf{1.00} \\ \hline
TRACE &
  \multicolumn{1}{c}{0.50} &
  \textbf{1.00} &
  \multicolumn{1}{c}{0.40} &
  \textbf{1.00} &
  \multicolumn{1}{c}{0.67} &
  \textbf{1.00} &
  \multicolumn{1}{c}{\textbf{1.00}} &
  \textbf{1.00} \\ \hline
In-Lab Arena &
  \multicolumn{1}{c}{0.56} &
  \textbf{1.00} &
  \multicolumn{1}{c}{0.33} &
  0.80 &
  \multicolumn{1}{c}{0.83} &
  \textbf{1.00} &
  \multicolumn{1}{c}{\textbf{1.00}} &
  \textbf{1.00} \\ \hline
NewlySim &
  \multicolumn{1}{c}{0.17} &
  \textbf{1.00} &
  \multicolumn{1}{c}{0.17} &
  \textbf{1.00} &  
  \multicolumn{1}{c}{0.17} &
  \textbf{1.00} &
  \multicolumn{1}{c}{\textbf{1.00}} &
  \textbf{1.00} \\ \hline
Open-World &
  \multicolumn{1}{c}{0.67} &
  0.80 &
  \multicolumn{1}{c}{0.38} &
  0.30 &  
  \multicolumn{1}{c}{0.35} &
  0.60 &
  \multicolumn{1}{c}{\textbf{0.91}} &
  \textbf{1.00} \\ \hline
\end{tabular}
\end{adjustbox}
\end{table}

\begin{table}[h]
\centering
\caption{The node-level accuracy results for \toolname and baselines. P stands for precision, and R stands for recall. }
\label{tab:node-level-cti}

\begin{adjustbox}{max width=0.96\linewidth}
\large % 或 \footnotesize 等，控制字体大小
\begin{tabular}{|c|cc|cc|cc|cc|}
\hline
 & \multicolumn{6}{c|}{\textbf{CTI Extraction Baselines}} & \multicolumn{2}{c|}{\textbf{Ours}} \\ \hline
\multirow{2}{*}{\textbf{Dataset}} &
  \multicolumn{2}{c|}{\begin{tabular}[c]{@{}c@{}}\textbf{\ttpdrill +}\\ \textbf{\toolname*}\end{tabular}} &
  \multicolumn{2}{c|}{\begin{tabular}[c]{@{}c@{}}\textbf{\ladder +}\\ \textbf{\toolname*}\end{tabular}} &
  \multicolumn{2}{c|}{\begin{tabular}[c]{@{}c@{}}\textbf{\toolname}\\  \textbf{with only IOC}\end{tabular}} &
  \multicolumn{2}{c|}{\textbf{\toolname}} \\ \cline{2-9} 
 &
  \multicolumn{1}{c}{P} &
  R &  
  \multicolumn{1}{c}{P} &
  R &
  \multicolumn{1}{c}{P} &
  R &
  \multicolumn{1}{c}{P} &
  R \\ \hline
THEIA &
  \multicolumn{1}{c}{0.31} &
  \textbf{1.00} &
  \multicolumn{1}{c}{0.15} &
  0.54 &
  \multicolumn{1}{c}{0.23} &
  0.88 &
  \multicolumn{1}{c}{\textbf{0.62}} &
  \textbf{1.00} \\ \hline
TRACE &
  \multicolumn{1}{c}{0.67} &
  \textbf{0.98} &
  \multicolumn{1}{c}{0.12} &
  0.63 &
  \multicolumn{1}{c}{0.18} &
  0.83 &
  \multicolumn{1}{c}{\textbf{0.82}} &
  \textbf{0.98} \\ \hline
In-Lab Arena &
  \multicolumn{1}{c}{0.54} &
  0.92 &
  \multicolumn{1}{c}{0.21} &
  0.52 &
  \multicolumn{1}{c}{0.32} &
  0.65 &
  \multicolumn{1}{c}{\textbf{0.82}} &
  \textbf{1.00} \\ \hline
NewlySim &
  \multicolumn{1}{c}{0.12} &
  0.84 &
  \multicolumn{1}{c}{0.08} &
  0.12 &
  \multicolumn{1}{c}{0.19} &
  0.68 &
  \multicolumn{1}{c}{\textbf{0.78}} &
  \textbf{1.00} \\ \hline
Open-World &
  \multicolumn{1}{c}{0.59} &
  0.37 &
  \multicolumn{1}{c}{0.11} &
  0.37&
  \multicolumn{1}{c}{0.11} &
  0.27 &
  \multicolumn{1}{c}{\textbf{0.82}} &
  \textbf{1.00} \\ \hline
\end{tabular}
\end{adjustbox}
\end{table}

For the ``\ttpdrill + \toolname'' and ``\toolname with only \ac{ioc}'' baselines, the reliance on \ac{ioc} alone, coupled with the limited number of corresponding \acp{ioc} in CTI reports, is the primary cause of poor performance. This limitation leads to numerous missed attack-related events, resulting in a large number of false negatives. Furthermore, both baselines generate many false positives, as \acp{ioc} are tied to specific paths, files, and command lines that don't necessarily indicate malicious behavior across all operations involving these entities.
In our evaluation, we observe that both baselines can only detect malicious external IP nodes and nodes with specific command lines or file names (e.g., the \textit{del} command used to erase attack traces, or the \textit{/etc/passwd} file accessed by attackers). However, they struggle to detect malicious processes or files disguised as normal nodes. Additionally, certain \ac{ioc}-related files and folders can also be accessed by legitimate processes. Relying on \ac{ioc} makes it difficult for \toolname to distinguish between these benign external IPs and legitimate command operations, leading to a significant number of false positives.

``\ladder + \toolname'' performs poorly because part of the extracted knowledge contains a significant amount of irrelevant information, which distorts the matching process with system events. More specifically, the incomplete and imprecise extraction by \ladder leads to incorrect query scores and ultimately disrupts the accurate scores derived from true \acp{ioc}.

In contrast, with \acp{gioc}, \toolname can identify attack behaviors, such as downloading of a malicious file from an external IP, the execution of the malicious file, and the removal of the malicious file after execution is completed, even if the names of malicious nodes don't appear in the \ac{cti}. This is because, using the event semantic lifting method in Section~\ref{sec:query}, \toolname can detect the malicious process and file node from the granularity of behavior and maps the \ac{gioc} dealing with the situation when the malicious file is renamed to disguise itself as a normal file.

\subsection{RQ 2: Accuracy of Technique Labeling}
\label{rq2:labeling}
A key advantage of \toolname is its ability to link ATT\&CK technique labels to low-level system events, yielding interpretable and precise detection.
With properly assigned ATT\&CK technique labels, \toolname can map system events to lifecycle stages. This capability enables temporal reasoning to filter out likely benign sequences, thereby enhancing detection precision without compromising interpretability. 

Therefore, in this section, we evaluates \toolname’s accuracy in assigning technique labels. The main challenge lies in establishing ground truth. While the Open-World dataset is fully labeled with ATT\&CK techniques, the In-Lab Arena dataset is partially labeled, and for the remaining unlabeled datasets,  we engaged the same red team in Open-World dataset to annotate techniques based on attack documents. To ensure consistency, each attack behavior was assigned at most three techniques during both annotation and expert review. A labeling is considered correct if \toolname matches any ground-truth techniques. Our evaluation covers 643 attack actions spanning 12 tactics and 65 techniques.

In our experiment, \toolname accurately labels 559 of all 643 attack actions, accounting for a 87.0\% accuracy. After manual inspection, 507 out of 559 actions can be labeled by \acp{gioc} while only 101 actions can also be identified by \acp{ioc}. For example, in the dataset, there is an event accessing the process memory file, (\textit{cp, read, /proc/11793/mem}), which is triggered by the attacker performing an OS credential dumping. For this event, no \ac{ioc} mentioned in \ac{cti} can be matched with the exact path of ``/proc/11793/mem''. However, there are related sentences in \ac{cti} stating ``The attackers usually dump the process memory in the system.'' With these sentences, \toolname can extract \ac{gioc} as (attacker, dump, process memory in the system), which can then be matched to the event. Of the 84 mislabeled actions, 69 are those that are not labeled by \toolname as the top 3 accurate labels, and the other 15 are incorrectly labeled by \toolname. 

The reason why \toolname does not give accurate label to the 69 attack actions is that these actions correspond to multiple attack techniques, causing ambiguity. For example, in the In-Lab Arena dataset, attackers downloaded malicious payloads by abusing PowerShell commands. This top 3 actions can correspond to \textit{Ingress Tool Transfer (T1105)}, \textit{Scheduled Task/Job (T1053)}, and \textit{Develop Capabilities-Malware (T1587.001)}, while \toolname gives \textit{Abuse PowerShell (T1059.001)}, which results in an incorrect label. Inspecting the candidate label set of this attack action in \toolname, we find that \textit{Ingress Tool Transfer (T1105)} is also a candidate, but with a lower similarity score.

For the other 15 mislabeled attack actions, it is because \toolname did not properly understand the semantics of the attack action, resulting in incorrect labeling. For example, in our datasets, the attackers used the \textit{echo} command to write invalid data to a malicious payload (execution file) to achieve an obfuscated operation, which reflected \textit{Obfuscated Files or Information (T1027)}. The \toolname incorrectly labeled this attack action as \textit{Windows Command Shell (T1059.003)} because it didn't understand the attack correctly and thought it was a simple malicious command execution.

Although labeling inaccuracies may appear concerning, they do not severely hinder attack reasoning, as predicted techniques are often semantically related to the ground truth. This enables \toolname to map them to similar lifecycle stages, preserving reasoning flow and preventing premature pruning of valid paths, thus avoiding graph-level false positives. Meanwhile, Section~\ref{sec:component} explains how attack reasoning further mitigates event-level false positives from technique labeling.
In addition to the overall analysis presented above, Section~\ref{casestudy} provides concrete examples illustrating how technique labels are captured by matching system events with \acp{gioc}. We recognize that our evaluation datasets and covered attack techniques are limited in representing real-world complexity. 

\begin{figure}[h]
 \centering
     \includegraphics[width=0.86\linewidth]{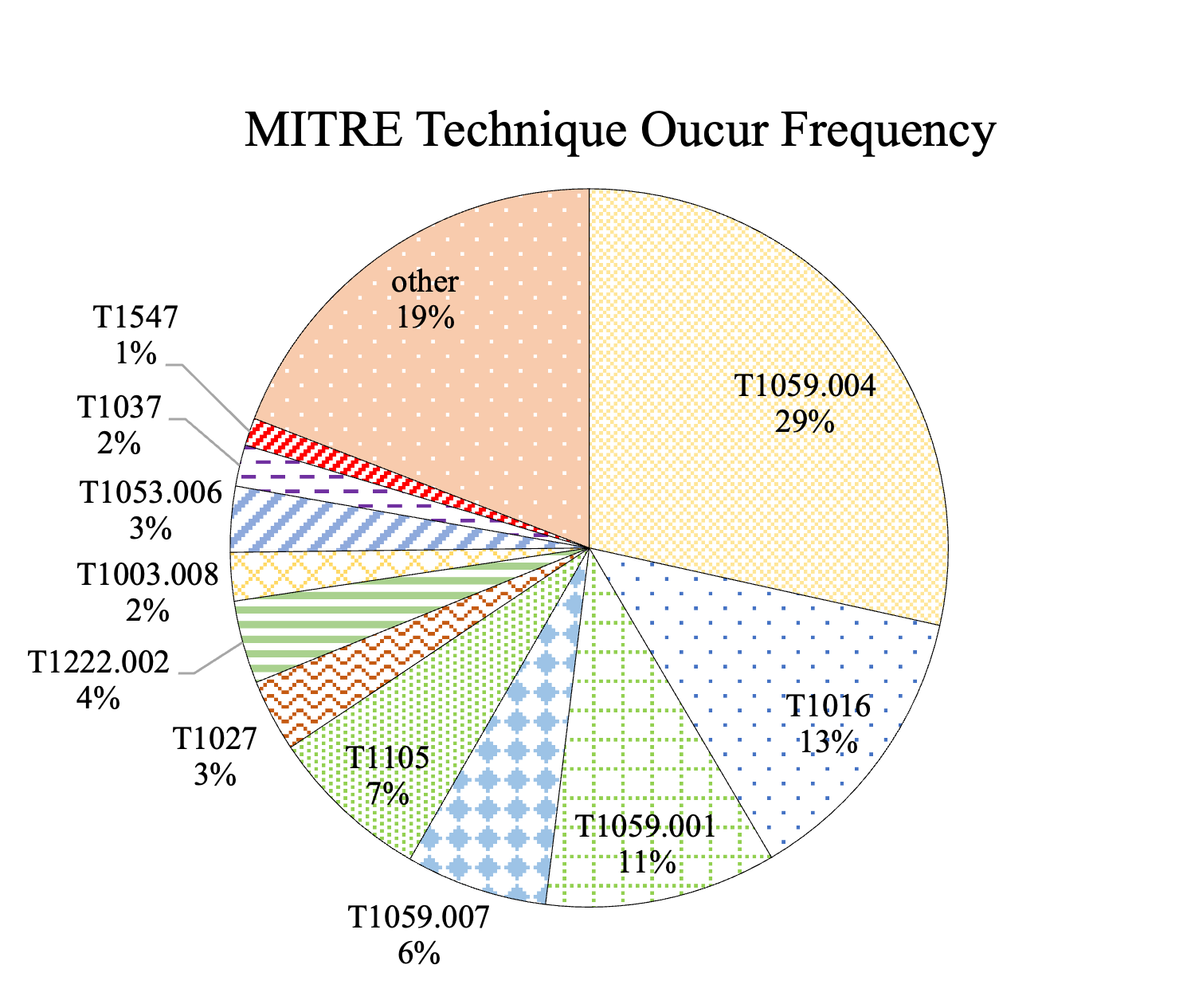}
 	\caption{Top 10 frequent techniques in our experiments.}
 	\label{fig:frequency}
 \end{figure}

To facilitate future research, Figure~\ref{fig:frequency} presents the ten most frequently observed MITRE ATT\&CK techniques in our experiments.  
The most common techniques are primarily associated with four tatics, including \textit{Execution}, \textit{Internal Reconnaissance}, \textit{Command and Control}, and \textit{Credential Access}. 
Among these, techniques under \textit{Internal Reconnaissance} (e.g., commands such as \textit{ifconfig} and \textit{arp}) are relatively easy to detect using \acp{ioc}, as often involving well-known system utilities. In contrast, the remaining techniques are more challenging to identify with \acp{ioc} alone, and require \acp{gioc} for accurate mapping. The case-level details are illustrated in Section~\ref{casestudy}.

\subsection{RQ 3: Efficiency}
To evaluate the efficiency, we compare the throughput of \toolname with baselines on different datasets. 
The throughput is defined as the number of system events processed per second (eps). We omit \ttpdrill and \ladder because they share the detection component of \toolname, leading to the same throughput.

\begin{figure}[t]
    \centering   
    \includegraphics[width=0.96\linewidth]{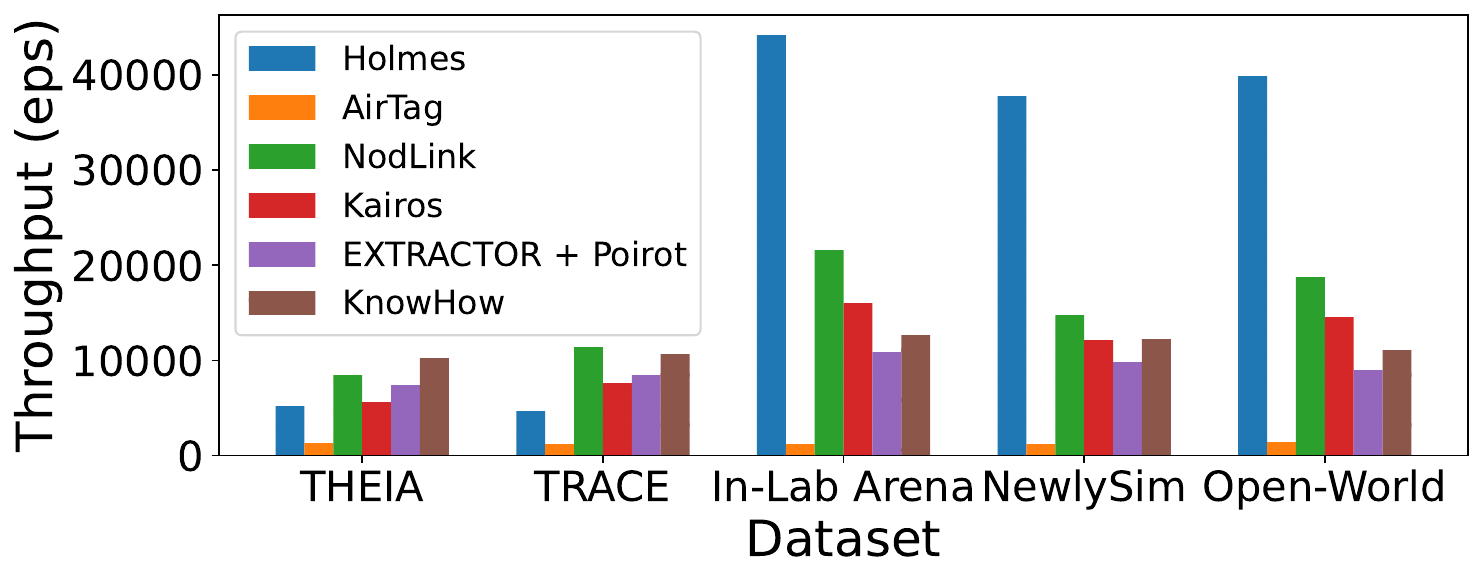}
	\caption{Working throughput among different frameworks. \ttpdrill and \ladder have the same throughput as \toolname.}
	\label{fig:throughput}
\end{figure}

Figure~\ref{fig:throughput} depicts the throughput of all the systems across different datasets. \toolname is comparable to the SOTA online detection works. \holmes achieves exceptionally high throughput in the In-Lab Arena and NewlySim datasets as it only has 16 rules, resulting in the lowest accuracy in Section~\ref{sec:accuracy}.
\airtag exhibits the lowest throughput because it uses the time-consuming BERT for event encoding.

Moreover, we analysis the impact of \ac{cti} scale on detection efficiency. Larger volumes of \acp{cti} may generate more \acp{gioc} within \ac{amid}, potentially slowing down $ProvQ$ performance. However, the theoretical time complexity of $ProvQ$ is $O(\log(n))$, where $n$ denotes the number of \acp{gioc}, and that is a complexity that remains acceptable even as the number of \acp{gioc} increases. Specifically, under the low-dimensional manifold hypothesis, which is empirically effective for textual data\cite{6472238, wang2021infobert}, our Mean-Shift algorithm clusters the $n$ \acp{gioc} into $k$ clusters, where $k=O(\log(n))$. Furthermore, by organizing the $k$ \ac{atie} clusters in \ac{amid} using a KD-tree structure, the approximate time complexity of $ProvQ$ becomes $O(k+\log(n/k))$\cite{10.1145/361002.361007, 10.1145/355744.355745}. Substituting $k=O(\log(n))$, this complexity simplifies to $O(\log(n))$. Therefore, the efficiency of \toolname remains largely unaffected by the scale of \ac{cti}.

\subsection{RQ 4: Robustness}
\label{sec:eva-robustness}

\noindent \textbf{Mimicry Attacks.}
Existing research has demonstrated that existing APT detection systems, particularly those based on graph learning, exhibit considerable vulnerability when facing mimicry attacks~\cite{goyal2023sometimes}. 
They find that attackers can mislead graph-learning-based detection approaches by inserting some benign behaviors into their attacks, causing the attack provenance graph to contain benign structures. The proportion of the size of the above-mentioned benign subgraph to the average size of normal benign structures in the benign data is called the insertion ratio. It shows that when the insertion ratio exceeds 2.00\%, the attacker can escape the detection algorithm.

To evaluate the robustness of \toolname against mimicry attacks, we constructed a mimicry dataset, Mimic-Prov, following the steps described in~\cite{goyal2023sometimes}, and evaluate \toolname on it. We include this dataset in our code repository. Note that we did not utilize th dataset openly sourced by~\cite{goyal2023sometimes} because it removes contextual information (\eg process names and command-line arguments) which is crucial for \toolname's detection. To accommodate the impact of insertion ratios of benign substructures on the system's performance, we constructed multiple datasets with four different ratios: 0.20\%, 0.50\%, 1.00\% and 2.00\%. Table~\ref{tab:robust} shows statistic details.

Our experimental results show that \toolname successfully detected ALL attack events at the graph level across these insertion ratios within the Mimic-Prov dataset. Notably, at a 1.00\% insertion ratio, node-level recall drops below 1, indicating successful evasion of some attack actions by mimic structures. Upon further analysis, we find that the missed nodes correspond to simple, shallow reconnaissance activities (e.g., ifconfig, arp), which lack complex event sequences and derived successor nodes, hence are more vulnerable to mimicry attacks. However, more complex and deep attacks, which dominate our datasets, are less susceptible to such interference.

\begin{table}[]
\centering
\caption{Detection result of \toolname on the Mimic-Prov dataset with different insertion ratios. Columns with an insertion ratio of 0.00 indicate the original dataset and no mimic structure is inserted. P, R are the same as Table~\ref{tab:graph-level}.}
\label{tab:robust}
\begin{adjustbox}{max width=\linewidth}
\normalsize % 或 \footnotesize 等，控制字体大小
\begin{tabular}{|c|cc|cc|cc|cc|cc|}
\hline
Ratio & \multicolumn{2}{c|}{0.00}        & \multicolumn{2}{c|}{0.20}        & \multicolumn{2}{c|}{0.50}        & \multicolumn{2}{c|}{1.00}        & \multicolumn{2}{c|}{2.00}        \\ \hline
Metrics         & \multicolumn{1}{c|}{P}    & R    & \multicolumn{1}{c|}{P}    & R    & \multicolumn{1}{c|}{P}    & R    & \multicolumn{1}{c|}{P}    & R    & \multicolumn{1}{c|}{P}    & R    \\ \hline
Graph           & \multicolumn{1}{c|}{1.00} & 1.00 & \multicolumn{1}{c|}{1.00} & 1.00 & \multicolumn{1}{c|}{1.00} & 1.00 & \multicolumn{1}{c|}{1.00} & 1.00 & \multicolumn{1}{c|}{1.00} & 1.00  \\ \hline
Node            & \multicolumn{1}{c|}{0.78} & 1.00 & \multicolumn{1}{c|}{0.43} & 1.00 & \multicolumn{1}{c|}{0.51} & 1.00 & \multicolumn{1}{c|}{0.31} & 0.90 & \multicolumn{1}{c|}{0.42} & 1.00 \\ \hline
\end{tabular}
\end{adjustbox}
\end{table}

\noindent \textbf{Incomplete Attacks.} We also evaluate how incomplete attack lifecycles impact the detection effectiveness and robustness of \toolname. We conduct a controlled experiment based on the In-lab Arena dataset. The In-lab Arena dataset was chosen for its unparalleled complexity of attack steps, enabling reliable incomplete-scenario construction. We simulate incomplete attacks by removing attack steps from the original attack sequences in the dataset. We define the attack integrity ratio as the ratio of remaining to original attack steps, and evaluate \toolname across the modified datasets with integrity ratios of 0.60, 0.70, 0.80, 0.90, and 1.00. Detection performance is measured in terms of precision and recall as in Section~\ref{protocol}.

The results are in Table~\ref{tab:integrity}. We observe that when the attack integrity ratio is 0.80 or higher, both graph- and node-level detection performance remain stable and effective, except for a limited increase in node-level false positives. It means removing a small portion of attack behaviors does not compromise the overall integrity of the attack lifecycle, thereby allowing normal alerting through attack reasoning.

When the attack integrity ratio falls below 0.80, we notice an increase in false negatives at the graph-level detection. For ratios of 0.60 and 0.70, two of three attack graphs are detected, while a simpler graph (less than 50 nodes) is missed. We further find that the missed attack graph only contains \textit{the Initial Compromise stage and Establish Foothold stage}, which is at the earlier stage of the attack. These stages have minimal behavioral distinction from benign activity, thus challenging for all knowledge-based methods.

\noindent\textbf{New and Unseen Attacks.} 
To further evaluate \toolname's capability in detecting unseen attacks, besides the statistic analysis on NewlySim dataset in Section~\ref{sec:accuracy}, we conducted a case study on the attack in NewlySim, which exploits CVE-2023-22809, a privilege escalation vulnerability absent from \ac{amid}, as one of attack actions, in Section~\ref{casestudy}.

\begin{table}[]
\centering
\caption{Detection result of \toolname on the incomplete attacks with different integrity ratios. Columns with the attack integrity ratio of 1.00 indicate the original dataset and no attack behaviors is removed. P, R are the same as Table~\ref{tab:graph-level}.}
\label{tab:integrity}
\begin{adjustbox}{max width=\linewidth}
\normalsize 
\begin{tabular}{|c|cc|cc|cc|cc|cc|}
\hline
Ratio & \multicolumn{2}{c|}{0.60}        & \multicolumn{2}{c|}{0.70}        & \multicolumn{2}{c|}{0.80}        & \multicolumn{2}{c|}{0.90}        & \multicolumn{2}{c|}{1.00}        \\ \hline
Metrics         & \multicolumn{1}{c|}{P}    & R    & \multicolumn{1}{c|}{P}    & R    & \multicolumn{1}{c|}{P}    & R    & \multicolumn{1}{c|}{P}    & R    & \multicolumn{1}{c|}{P}    & R    \\ \hline
Graph           & \multicolumn{1}{c|}{1.00} & 0.67 & \multicolumn{1}{c|}{1.00} & 0.67 & \multicolumn{1}{c|}{1.00} & 1.00 & \multicolumn{1}{c|}{1.00} & 1.00 & \multicolumn{1}{c|}{1.00} & 1.00 \\ \hline
Node            & \multicolumn{1}{c|}{0.76} & 0.83 & \multicolumn{1}{c|}{0.76} & 0.82 & \multicolumn{1}{c|}{0.84} & 1.00 & \multicolumn{1}{c|}{0.74} & 1.00 & \multicolumn{1}{c|}{0.82} & 1.00 \\ \hline
\end{tabular}
\end{adjustbox}
\end{table}

\subsection{RQ 5: Component-wise Analysis}
\label{sec:component}
We conduct a component-wise analysis to illustrate the impact of each component of \toolname by measuring node-level precision and recall after each step. We follow the same evalution protocal mentioned in Section~\ref{protocol}. The steps includes event-level detection with gIoC (Step 1), constructing provenance graphs based on anomalous events (Step 2) and attack reasoning (Step 3). This breakdown reveals how successive stages refine results, balancing early-stage trade-offs. The results are in Table~\ref{tab:component-rq} and we analyze it as follows.

\begin{table}[]
\caption{The intermediate detection result for each component of \toolname. P, R are the same as Table~\ref{tab:graph-level}.}
\resizebox{\columnwidth}{!}{%
\begin{tabular}{|c|cc|cc|cc|}
\hline
\multirow{2}{*}{\begin{tabular}[c]{@{}c@{}}Node-level   \\ Result\end{tabular}} & \multicolumn{2}{c|}{Event-level Detection} & \multicolumn{2}{c|}{Graph Construction} & \multicolumn{2}{c|}{Attack Reasoning} \\ \cline{2-7} 
                                                                                & \multicolumn{1}{c|}{P}             & R             & \multicolumn{1}{c|}{P}            & R           & \multicolumn{1}{c|}{P}           & R           \\ \hline
THEIA                                                                           & \multicolumn{1}{c|}{0.47}          & 0.74          & \multicolumn{1}{c|}{0.41}         & 1.00        & \multicolumn{1}{c|}{0.62}        & 1.00        \\ \hline
TRACE                                                                           & \multicolumn{1}{c|}{0.67}          & 0.83          & \multicolumn{1}{c|}{0.60}         & 0.98        & \multicolumn{1}{c|}{0.82}        & 0.98        \\ \hline
In-Lab Arena                                                                    & \multicolumn{1}{c|}{0.52}          & 0.74          & \multicolumn{1}{c|}{0.46}         & 1.00        & \multicolumn{1}{c|}{0.82}        & 1.00        \\ \hline
NewlySim                                                                        & \multicolumn{1}{c|}{0.48}          & 0.70          & \multicolumn{1}{c|}{0.46}         & 1.00        & \multicolumn{1}{c|}{0.78}        & 1.00        \\ \hline
Open-World                                                                      & \multicolumn{1}{c|}{0.63}          & 0.79          & \multicolumn{1}{c|}{0.57}         & 1.00        & \multicolumn{1}{c|}{0.82}        & 1.00        \\ \hline
\end{tabular}%
}
\label{tab:component-rq}
\end{table}

Step 1 initially identifies high-confidence anomalies (e.g., Lua script execution, sudo privilege escalation). However, events with benign-like features (e.g., ambiguous network connections) may fall below the detection threshold, limiting recall (In-Lab Arena: 74\% recall).
Step 2 improves recall (up 26\%) by propagating scores to connected nodes. For example, undetected attacker-controlled sshd nodes inherit scores from detected downstream actions and these scores then exceed the detection threshold, enabling the previously missed nodes to be identified. However, over-propagation to benign nodes (e.g., hostguard spawning malicious sh) introduces false positives, reducing precision (down 6\%).
Step 3 restores precision (up 36\%) by enforcing lifecycle constraints. Temporally inconsistent nodes (e.g., hostguard process incorrectly labeled in Establish Foothold) are filtered, while fragmented attacks are reconstructed through subgraph alignment which improves the final node-level precision.

In  conclusion, our three-stage design systematically addresses earlier limitations: propagation recovers missed signals (recall), while knowledge-guided reasoning suppresses contextual false positives (precision). The interplay demonstrates \toolname's robustness against sophisticated APT attacks.

\eat{
Fortunately, despite the presence of false negatives under these conditions, if we abandon the attack completeness assumption and instead use the $\sigma$-based-outlier detection method based on attack lifecycle scores , all attacks are detected for integrity ratios of 0.50, 0.60, and 0.70, with only 3, 3, and 2 benign lifecycles being falsely flagged as attacks, respectively. Although this approach sacrifices some visualization effectiveness due to the lack of attack completeness assumptions, it achieves zero false negatives with relatively low false positive rates.

Therefore, we acknowledge that incomplete attacks may interfere with \toolname's detection mechanism, potentially allowing certain portions of the attack lifecycles to evade identification. However, such incomplete attack activities typically occur at a very early stage and tend to cause minimal damage to the victim system. As soon as attackers perform additional malicious actions to improve the integrity of their attacks, \toolname is able to accurately detect and flag the corresponding activities. Moreover, our experiments demonstrate that by relaxing the assumption of attack completeness, \toolname can still achieve the full attack detection at a relatively low false positive rate.}

\section{Case Study}\label{sec:case-study}
\label{casestudy}

In this section, we analyze how \toolname detects unseen attacks and \ac{lotl} techniques through case studies on the NewlySim and Open-World datasets.

\begin{figure}[h]
     \includegraphics[width=0.96\linewidth]{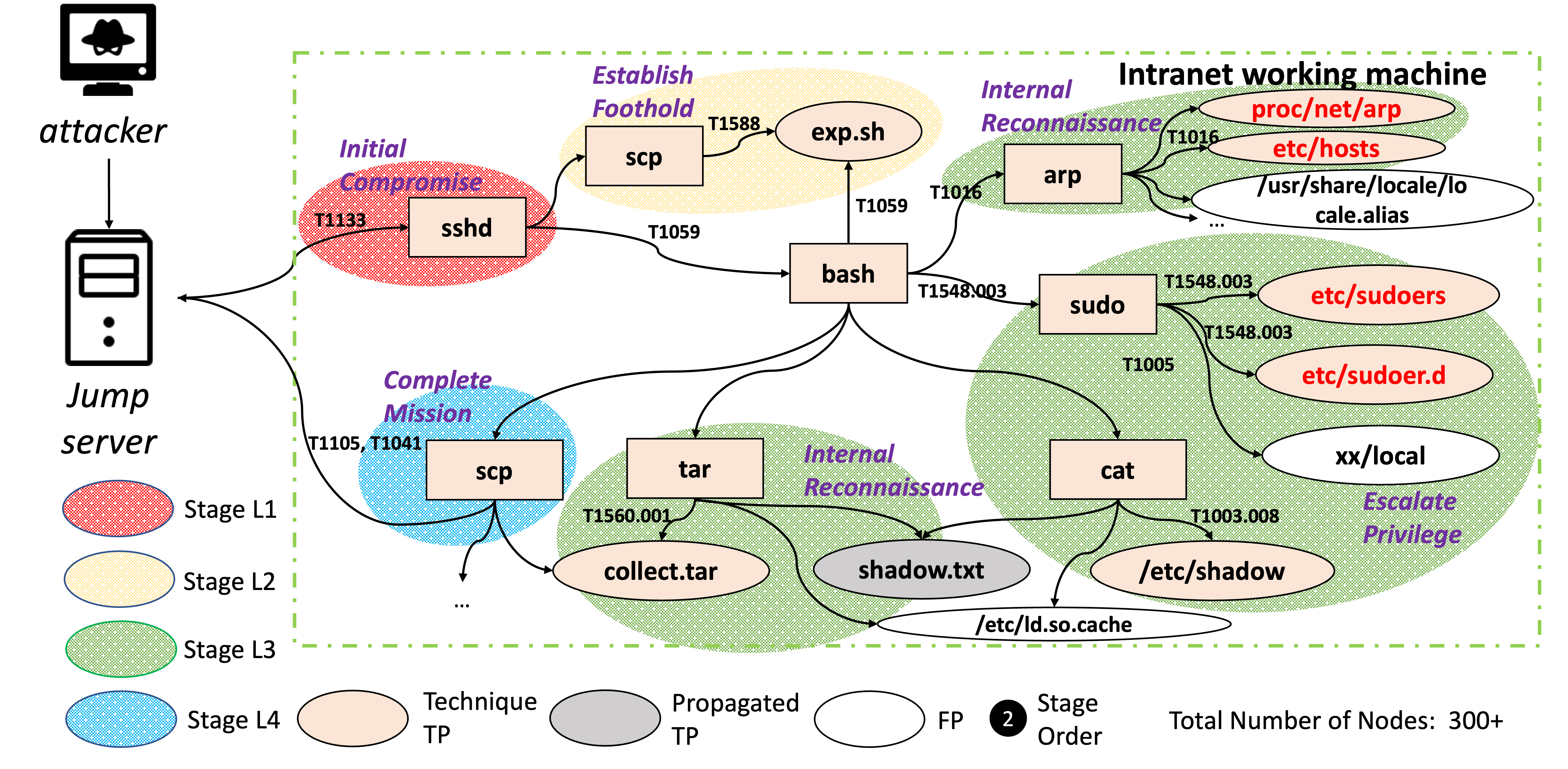}
 	\caption{Detection result of the APT campaingn using CVE-2023-22809 from the NewlySim dataset.}
 	\label{fig:casestudy1}
 \end{figure}

\textbf{NewlySim Dataset. }Figure~\ref{fig:casestudy1} shows the detection result of \toolname on an APT campaign from NewlySim dataset, which exploits CVE-2023-22809—a privilege escalation vulnerability that not documented in \ac{amid}. Attackers first uploaded the exploit script \textit{exp.sh} via \textit{scp}, then executed it to gain \textit{sudo} privileges through \textit{sudoedit}, accessing critical system files. Finally, attackers collected and compressed password files and transferred them to the jump server using \ac{lotl} processes (\textit{cp}, \textit{tar}, \textit{scp}).

Despite CVE-2023-22809 being undocumented in \ac{amid} and the \ac{lotl} actions, \toolname successfully detected all attack actions by attackers through \acp{gioc} with few false positives, whereas the SOTA baselines failed. This capability stems from three key factors: 
\ding{172} Although CVE-2023-22809 is not documented, \ac{amid} contains similar attack behaviors. For instance, the \ac{atie} of T1548.003 includes \acp{gioc} like (Adversary, read, sudoers file), (Adversary, perform, sudoers file), which align with system events generated when accessing the \textit{sudoers} file. \toolname uses this semantic similarity to detect novel vulnerabilities and infer attack intent.
\ding{173} \toolname can identifies \ac{lotl} behaviors involving legitimate tools like \textit{cat}, \textit{scp}, and \textit{tar}. For example, the \ac{atie} of T1548.003 contains the \ac{gioc}, (Adversary, read, etc shadow file), matching events where the attacker used \textit{cp} to read the \textit{shadow} file. Event Semantic Lifting and Event and gIoC Embedding first lift the ``cat" command to ``show", then reduce the distance between ``read" and ``show" through embedding, allowing \toolname to realize the \ac{lotl} actions.
\ding{174} Attack reasoning enables \toolname to reduce false positives. For instance, other benign \textit{tar} operations of normal users accessing shared files (e.g., \texttt{/etc/ld.so.cache} in Figure~\ref{fig:casestudy1}) that are often misclassified by data-driven baselines, are correctly identified as non-malicious through contextual analysis in attack reasoning.

\begin{figure}[h]
     \includegraphics[width=0.96\linewidth]{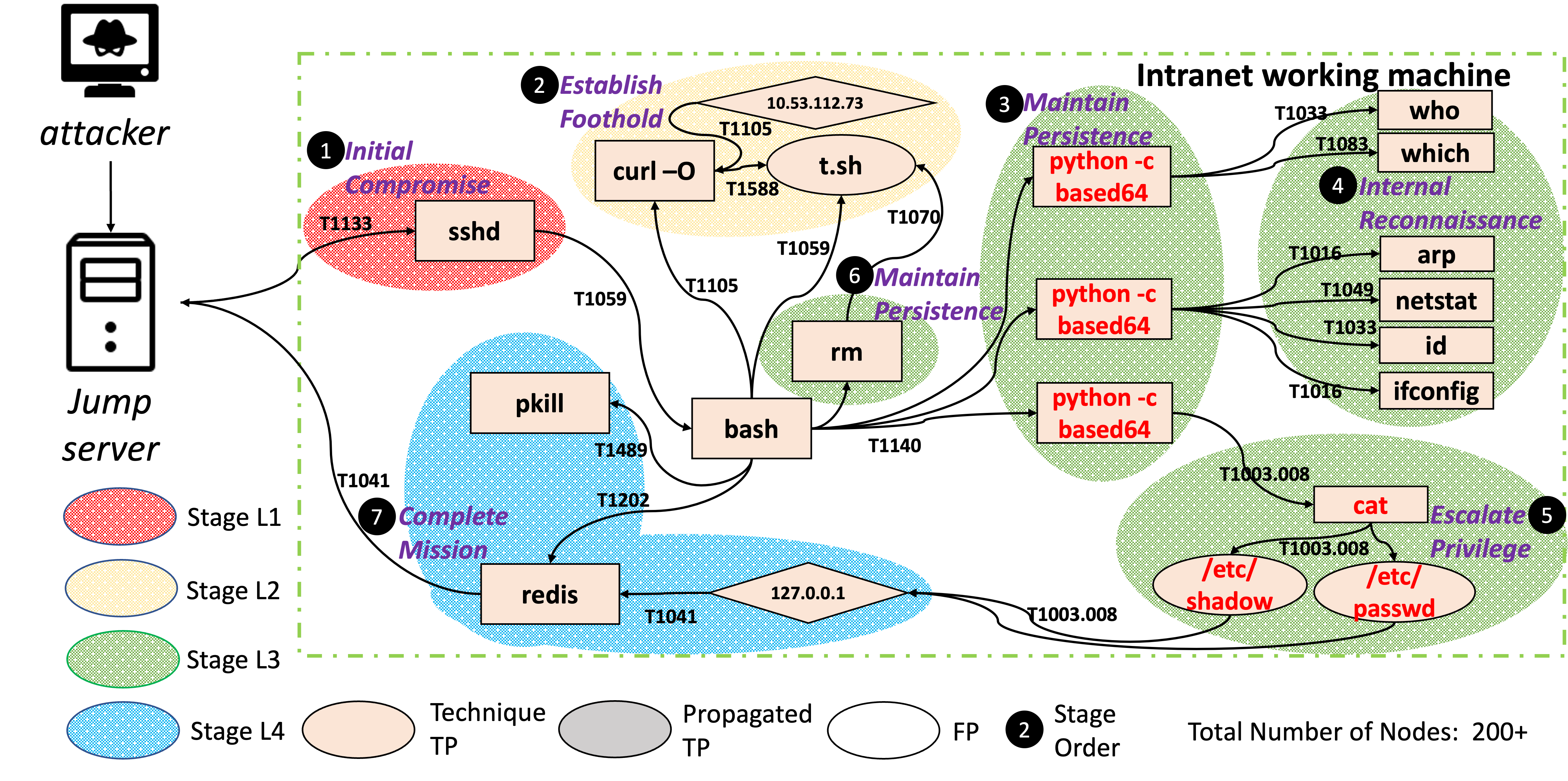}
 	\caption{Detection result of the APT case in the Open-World dataset.}
 	\label{fig:casestudy2}
 \end{figure}

\textbf{Open-World Dataset. } Figure~\ref{fig:casestudy2} presents the detection result of \toolname on an APT campaign from the Open-World dataset that exemplifies the extensive use of \ac{lotl} techniques\eat{. This attack campaign demonstrates a typical \ac{lotl} strategy where attackers rely primarily on legitimate system tools and commands}, leaving minimal forensic footprints. 
Specifically, attackers first accessed to the target host from a compromised server, and leveraged the \ac{lotl} command, \textit{curl}, to download a malicious script, \textit{t.sh}. Subsequently, attackers executed \textit{t.sh}, which invoked three Base64-encoded, obfuscated Python commands to perform internal network reconnaissance and sensitive data exfiltration. \eat{These commands performed internal network scanning and sensitive data exfiltration. }To evade detection, attackers used the \ac{lotl} command, \textit{rm}, to delete the script after execution. Finally, the stolen data is exfiltrated via the built-in \textit{redis} service, which is then terminated by the \ac{lotl} command, \textit{pkill}, to eliminate evidence of the attack.

Although this attack campaign employed several \ac{lotl} techniques, \toolname achieved full detection with minimal false positives, while the other baselines failed. This capability also stems from the same three key factors:
\ding{172} \ac{amid} contains the attack knowledge that  that closely reflects attacker behaviors. For example, the \ac{atie} of T1140 includes \acp{gioc} such as (Adversary, use, base64 encoded file) and (Attacker, run, base64 obfuscated scripts), which aligns with the attacker’s use of obfuscated Python commands in the event-level detection. Similarly, other malicious activities, such as downloading the \textit{t.sh} script via \textit{curl} from a compromised host, removing the script, exfiltrating sensitive data, and performing internal reconnaissance, were also successfully detected at the event level. \ding{173} \toolname can effectively identify \ac{lotl} behaviors. For example, the \ac{lotl} behaviors, such as invoking \textit{redis}, \textit{who}, and \textit{id}, are assigned high scores with the existing \ac{amid} knowledge are subsequently filtered out during event-level detection. \ding{174} Attack reasoning enables KNOWHOW to reduce false positives especially on the \ac{lotl} behaviors. For example, data-driven approaches generates numerous false positives in this case, because the \textit{redis} service is widely used legitimate system users, and its benign usage can easily be misclassified as malicious.
\toolname significantly reduces such false positives during the attack reasoning step. This is because this false positives occurred either before the Initial Access stage or after the Complete Mission stage and were therefore filtered out by \toolname, significantly reducing false alarms.

\section{Other Related Work}
\noindent\textbf{Data-Driven Provenance Analysis.}
Data-driven methods
~\cite{nodlink, shadewatcher,wang2020you,atlas,kairos,depcomm,airtag, magic, flash, rcaid} are widely studied in \ac{apt} detection. These methods typically employ deep learning models to extract features from provenance graphs and then apply various algorithms to classify nodes or subgraphs as benign or malicious. Different from CTI-based methods, data-driven methods do not rely on static knowledge of attack techniques and vulnerabilities but instead learn from historical data to identify patterns and anomalies. However, they often face challenges such as high resource consumption, high false positive rates, and difficulties in interpreting results~\cite{shaofeiempirical}.

\section{Discussion}

\noindent\textbf{Mimicry Attacks and Evasion.} 
A potential evasion involves an adaptive attack where the attacker, aware of \toolname's design, blends attack steps with crafted benign steps to mislead the one-to-limited mapping of reasoning module, as happened in Section~\ref{sec:eva-robustness}. For this to succeed, attackers must: \ding{172} effectively misclassify most attack steps using plausible benign actions, otherwise \toolname can infer the attack lifecycle from remaining steps; \ding{173} have full knowledge of \ac{amid} and \toolname’s training data, requiring extensive local testing to refine attack sequences. These conditions make such attacks technically challenging and require nontrivial work. Hence, while theoretically possible, the likelihood of a successful adaptive attack evading \toolname appears very low.

\noindent\textbf{Limitations}: 
We acknowledge three limitations in the design and evaluation of \toolname. First, while \toolname demonstrates scalability to previously unseen and \ac{lotl} attacks by capturing behavioral mechanisms (as demonstrated in Section~\ref{sec:case-study}), it cannot detect entirely novel attacks where every step is completely unprecedented and exhibits no similarity to known behaviors in the \ac{amid}. Although such cases are rare in practice, this limitation is inherent to all knowledge-driven detection approaches. Second, like prior systems (e.g., \nodlink, \holmes, \kairos), \toolname requires a complete attack graph for accurate alerts, which introduces detection delays and makes early-stage detection challenging due to limited information. However, the experimental results in Section~\ref{sec:eva-robustness} demonstrate that \toolname can still achieve accurate detection with relatively low false positive rates, even before the attack has fully completed. Third, our current evaluation is based on a limited set of CTI reports and datasets. These datasets cover only a subset of known attack techniques, as illustrated in Section~\ref{rq2:labeling}, and the types and distributions of these techniques may not fully reflect real-world attack scenarios. Therefore, whether \toolname can maintain strong performance in more complex, real-world environments remains to be validated through future deployment efforts.

\section{Conclusion}
In this paper, we propose \toolname, a \ac{cti}-knowledge-driven online provenance analysis solution that can automatically learn high-level attack knowledge from \ac{cti} reports and apply this knowledge to detect \ac{apt} attacks in low-level system events. 
Our evaluation shows that \toolname outperforms existing baselines in terms of both accuracy and interpretability. For interpretability, \toolname can automatically map system events to high-level technique descriptions and summarize them into APT Lifecycle stages, while none of the baselines can achieve this goal.  Furthermore, \toolname maintains the same level of efficiency as the baselines and is robust to attacks stemming from unknown vulnerabilities and existing mimicry attacks.

\section*{Acknowledgments}
This work was partly supported by the National Science and Technology Major Project of China (2022ZD0119103), the National Natural Science Foundation of China (62172009), and Huawai Research Fund.

%\clearpage
% \input{tex/Acknowledgments}

%for ACM papers
% \bibliographystyle{ACM-Reference-Format}
%for IEEE papers
% \input{tex/ethic}

\bibliographystyle{IEEEtranS} %for IEEE papers
\bibliography{references}

%\input{tex/Appendix}

% that's all folks
\end{document}